\documentclass[prb,aps,twocolumn,showpacs,amsmath,floatfix,superscriptaddress]{revtex4}

\usepackage{graphicx,epsf,natbib,amsmath,latexsym,amssymb}
\usepackage{color}
\def\be{\begin{equation}}
\def\ee{\end{equation}}
\def\bea{\begin{eqnarray}}
\def\eea{\end{eqnarray}}

\begin{document}
\title{Tunneling between helical Majorana modes and helical Luttinger liquids}

\author{Sung-Po Chao}
\affiliation{Institute of Physics, Academia Sinica, Taipei 11529, Taiwan, R.O.C.}
\affiliation{Physics Division, National Center for Theoretical Science, Hsinchu, 30013, Taiwan, R.O.C.}
\author{Thomas L.~Schmidt}
\affiliation{Department of Physics, University of Basel, Klingelbergstrasse 82, 4056 Basel, Switzerland}
\affiliation{Physics and Materials Science Research Unit,
University of Luxembourg, L-1511 Luxembourg}
\author{Chung-Hou Chung}
\affiliation{Physics Division, National Center for Theoretical Science, Hsinchu, 30013, Taiwan, R.O.C.}
\affiliation{Electrophysics Department, National Chiao-Tung University, Hsinchu, 30010, Taiwan, R.O.C.}
\date{\today}

\begin{abstract}
We propose and study the charge transport through single and double quantum point contacts setup between helical Majorana modes and an interacting helical Luttinger liquid. We show that the differential conductance decreases for stronger repulsive interactions and that the point contacts become insulating above a critical interaction strength. For a single point contact, the differential conductance as a function of bias voltage shows a series of peaks due to Andreev reflection of electrons in the Majorana modes. In the case of two point contacts, interference phenomena make the structure of the individual resonance peaks less universal and show modulations with different separation distance between the contacts. For small separation distance the overall features remain similar to the case of a single point contact.   
\end{abstract}
\maketitle

\section{Introduction}
The recent discovery of topological insulators\cite{Hasan,Qi} has spurred tremendous interest in the topological phases of condensed-matter systems. Topological systems in two-dimensional (2D) systems are characterized by their peculiar symmetry-protected gapless one-dimensional (1D) edge states in the presence of a gapped bulk.\cite{Essin,Mong} In time-reversal invariant (TRI) systems, two types of 1D edge states are especially remarkable:

On the one hand, helical Dirac fermions, whose spin is locked to the momentum, were first theoretically predicted\cite{Bernevig,KaneMele} and experimentally realized\cite{Konig,Knez} as the edge states of 2D topological insulators. Being rather insensitive to disorder, these edge states have promising applications in the fields of nanoelectronics and spintronics.

On the other hand, helical Majorana modes have been predicted to exist as the edge state of TRI topological superconductors.\cite{FuKane,XLQi,Tanaka,Sato,Liu,Wang,Queiroz} The current interest in the search for various Majorana modes\cite{CWJ,Alicea} in condensed-matter systems mainly stems from their possible applications in fault-tolerant quantum computing. While some experimental signatures for Majorana zero modes existing as the end states of effective 1D topological superconductors have already been found,\cite{Mourik,NadjPerge} conclusive evidence in particular of their non-Abelian exchange properties is still actively sought for. 

In this paper we focus on the charge transport between a system of 1D helical Dirac fermions and a system of helical Majorana modes, which are tunnel-coupled by one or several quantum point contacts. Due to the Coulomb interaction between the electrons, the low-energy properties of the helical Dirac fermions is described by the helical Luttinger liquid theory\cite{Wu,Xu} and is possibly realized in the InAs/GaSb experimental setup by Du's group\cite{Du}. On the other hand, the helical Majorana modes can to a good approximation be treated as free Majorana fermions. While strong interactions between the constituent electrons and holes\cite{Nagaosa} may destabilize the Majorana modes,\cite{Loss} the nearby superconductor screens moderate interactions effectively.\cite{Orth15} As long as the Majorana modes exist, they behave largely as chargeless particles and can be regarded as free.

Similar tunneling phenomena in heterostructures have been discussed using renormalization group (RG) analysis\cite{Fidkowski,YW,Affleck} and scattering formalism\cite{Fidkowski,Law,Li} in the case of noninteracting lead(s). In this paper, we calculate the tunneling current by using perturbation theory in the coupling between the Majorana modes and the interacting helical lead, using the interacting helical lead Green's function obtained by bosonization and the noninteracting Majorana Green's function as the unperturbed propagators. We use a scaling analysis to establish that the tunneling term is the most renormalization-group relevant local perturbation in our system. We consider a finite-size topological superconductor with discrete helical Majorana energy levels and assume for simplicity that the level separation is larger than the tunneling rate. With this assumption, we derive analytic results for the tunneling current through one or two quantum point contacts, and obtain the current-voltage relation by evaluating the analytic results numerically. 

For a single quantum point contact with a noninteracting lead, the tunneling current is the same (up to an extra factor of two in the differential conductance due to two spins) as for chiral Majorana fermions.\cite{Law} It shows periodic peak structures originating from the perfect Andreev reflection in different Majorana energy levels. With increasing repulsive interaction strengths, corresponding to a smaller Luttinger parameter $K$, the differential conductance begins to decrease and eventually vanishes completely at the resonance positions.
This effect may partly explain why perfect Andreev reflection is difficult to observe even despite the existence of Majorana zero modes in quantum wire experiments.\cite{Mourik} The quantum critical behavior for the tunneling through a single point contact is similar to the charge transport with two helical Luttinger leads connected by a quantum dot\cite{Schmidt,sp} or the information leakage in the helical lead connected to a Majorana mode.\cite{shHo}

As in a real experiment, the tunneling may not be perfectly local, we shall also consider the effect of extended point contacts\cite{Martin,Sassetti} using a model involving two point contacts. For two nearby quantum point contacts, the distance between the two contacts determines the interference structures in the tunneling current.\cite{Dolcini,Orth13} For distances much smaller than the boundary length of the topological superconductor, the interference changes the shape of the individual peaks, giving features similar to Fano resonances with the overall magnitude\cite{footnote} remaining periodic. When the separation distance is comparable with the boundary length we see the overall
magnitude also experiences some modulations related to the separation scale. Those interference features make the transport signature less universal and possibly modify the scaling behavior,\cite{Sassetti} constituting another reason why perfect transmission is hard to observe.

This paper is organized in the following way: in Sec.~II we present the setup, the corresponding model Hamiltonian, and the perturbation scheme. In Sec.~III we use a scaling analysis to identify the tunneling term as the most relevant term in our system and make a comparison with other systems or different boundary conditions. In Sec.~IV we present the main analytic and numerical results for single and double quantum point contacts, and discuss their physical interpretations. We summarize our results and compare our approach to using scattering eigenstates in Sec.~V.     
  
\section{Tunnel junctions}
\subsection{Proposed setup}

\begin{figure}
\includegraphics[width=1\columnwidth]{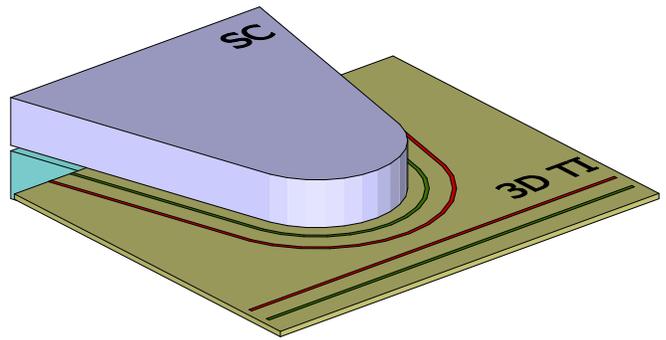}
\caption{Proposed setup for realizing heterojunction(s) of helical Luttinger lead and helical Majorana modes using thin film of three dimensional topological insulator. The top and bottom superconductors' order parameters carry different signs as proposed in the Ref.~\onlinecite{Liu}.}
 \label{setup}
\end{figure}

To realize both 1D helical Luttinger liquids as well as 1D helical Majorana modes, we adapt a proposal of Ref.~[\onlinecite{Liu}] based on thin films of 3D topological insulators (3DTI) such as $\text{Bi}_2\text{Se}_3$ or $\text{Bi}_2\text{Te}_3$, see Fig.~\ref{setup}. Helical Dirac fermions emerge at the sample edges due to the mixing of the top and bottom surface bands of the thin film.\cite{Zhang, Shen} Moreover, helical Majorana modes can be realized by sandwiching the thin film between conventional $s$-wave superconductors. For opposite signs of the superconductor pairing functions on the top and bottom layer, and sufficiently strong proximity-induced pairing amplitude (greater than the mixing gap of the two surfaces), helical Majorana modes are indeed formed as the edge states of the thin film.\cite{Liu}

\begin{figure}
\includegraphics[width=1\columnwidth]{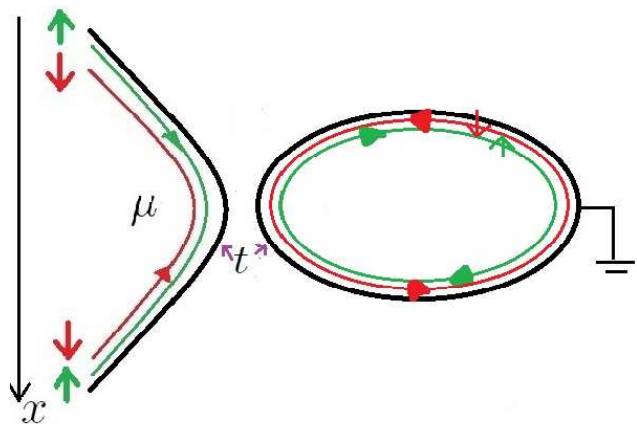}
\caption{Schematic figure of single tunneling junction}
 \label{set}
\end{figure}

In contrast to the converter between helical Dirac fermions and Majorana modes proposed in Refs.~[\onlinecite{Liu, Beri}], we study the charge transport through tunneling junction(s) between a helical Luttinger liquid and helical Majorana modes connected to ground. For systems with broken time-reversal symmetry, where the helical edge states are replaced by chiral ones, such transport phenomena were studied with noninteracting leads\cite{Law} and interacting leads.\cite{YW} The schematic diagram for single tunneling junction is depicted in Fig.~\ref{set}, where the voltage difference between the two leads is controlled by the chemical
potential $\mu$ imposed on the helical Luttinger liquid lead. The tunneling amplitude $\bar{t}$ is controlled by the width of the junction and is related to the wavefunction overlap between the two leads.  
 
In the real experiments it may not be easy to fix the relative phases of two adjacent superconductors nor fine tune the chemical potential to the topological regimes. For the heterostructure setup we can make the helical Majorana modes by different types of realizations\cite{realhm}, or change the helical modes to
time reversal preserved double Majorana end states\cite{Law2}. The results for tunnel transport to double Majorana end states is the limiting case of helical Majora mode with energy level separations going to infinity, as is shown in Fig. \ref{f4}.

\subsection{Model Hamiltonian}
\begin{figure}
\includegraphics[width=1\columnwidth]{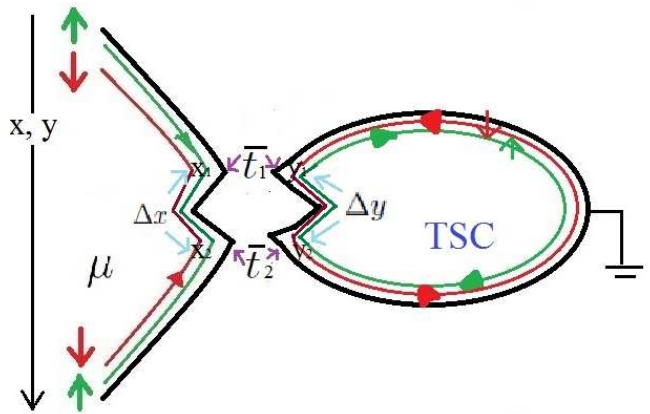}
\caption{Schematic figure for two tunneling junctions separated by spatial distance $\Delta x$ and $\Delta y$}
 \label{setn2}
\end{figure}

We consider one helical Luttinger liquid lead and one helical Majorana fermion lead. The Hamiltonian describing this system is 
$H= H_{L}+H_{M_0}+\sum_\alpha H_{T_\alpha}+\delta H$. The Hamiltonian $H_L = \int_{-\infty}^\infty dx \mathcal{H}_L$ for the helical fermions is a Luttinger liquid Hamiltonian density,
\begin{eqnarray}\label{eq1}
&&\mathcal{H}_L=iv_F\left[\psi_{L}^{\dagger}(x)\partial_x \psi_{L}(x)-\psi_{R}^{\dagger}(x)\partial_x \psi_{R}(x)\right]\\\nonumber
&&-\mu(x) \left[\psi_{L}^{\dagger}(x)\psi_{L}(x)+\psi_{R}^{\dagger}(x)\psi_{R}(x)\right]\\\nonumber
&&+u_2\psi_{L}^{\dagger}(x)\psi_{L}(x)\psi_{R}^{\dagger}(x)\psi_{R}(x)\\\nonumber
&&+\sum_{r=R,L}\frac{u_4}{2}\psi_{r}^{\dagger}(x)\psi_{r}(x)\psi_{r}^{\dagger}(x)\psi_{r}(x)\\\nonumber
\end{eqnarray}
The Hamiltonian for the grounded propagating Majorana fermions on a ring of circumference $L$ is
\begin{eqnarray}
H_{M_0}=i\sum_{\sigma}\int_0^L dx \left(v_{M,\sigma}\gamma_{\sigma}(x)\partial_x \gamma_{\sigma}(x)\right)
\end{eqnarray}
Here $v_{M,\sigma}=\text{sgn}(\sigma)v_M$.
The single particle tunneling term between the helical Luttinger liquid lead and helical Majorana fermion lead is described by\cite{Law}
\begin{eqnarray}
H_{T}=i\sum_{r,\sigma,\alpha}\frac{t_{r\sigma\alpha}}{\sqrt{2}}\gamma_{\sigma}(y_\alpha)\left[\xi_{r\sigma\alpha}\psi_{r}(x_\alpha)+\xi_{r\sigma\alpha}^{\ast}\psi_{r}^{\dagger}(x_\alpha)\right]
\end{eqnarray}
Here, $t_{r\sigma\alpha}$ is the tunneling strength, and $\xi_{r\sigma\alpha}$ are complex numbers with $|\xi_{r\sigma\alpha}|=1$. $r$ indicates the left/right movers in $H_L$, $\sigma$ denotes the spin index of the Majorana fermions, and $\alpha=1,..,N$ is the number of tunneling channels (junctions) and $x_\alpha/ y_\alpha$ are their spatial coordinates in Luttinger/Majorana leads. We restrict our discussions to $N=1$ and $N=2$ in this paper but the extension to arbitrary $N$ is straightforward and similar to the $N=2$ case. The $N=2$ case is illustrated in Fig.~\ref{setn2}.
 
The remaining $\delta H$ term contains the leading instabilities\cite{Fidkowski,YW} under the renormalization group analysis in the low-energy sector. We show in the next section why they are not important in our setup. With this simplification the full low-energy effective Hamiltonian becomes $H\simeq H_{L}+H_{M_0}+\sum_{\alpha}H_{T_\alpha}$, describing the single-particle tunneling between spinful Luttinger liquids and Majorana fermions lead. 
The tunneling charge current through site $x_\alpha$ (in the helical Luttinger liquid coordinate system) is obtained by 
\begin{eqnarray*}
\langle \hat{I}_{x_\alpha}\rangle&=&ie\langle[\sum_{r}\psi_{r}^{\dagger}(x)\psi_{r}(x),H]\rangle\\
&=&-e\langle\sum_{r,\sigma}\frac{t_{r\sigma\alpha}}{\sqrt{2}}\gamma_{\sigma}(y_\alpha)\left(\xi_{r\sigma\alpha}\psi_{r}(x_\alpha)-\xi_{r\sigma\alpha}^{\ast}\psi_{r}^{\dagger}(x_\alpha)\right)\rangle
\end{eqnarray*}

The total tunneling current is the coherent sum of the current from all tunneling channels (under the assumption that the separation distance between junctions are less than the coherence length). We choose a time dependent gauge transformation to move the chemical potentials in $H_{L}$
to $H_{T}$ by writing $\psi_{R/L}\rightarrow e^{i\mu t}\psi_{R/L}$. 
By defining the Keldysh contour ordered Green's function $G_{\sigma,R/L,\alpha}(t,t')=-i\langle T_c\{\gamma_\sigma(y_\alpha,t)\psi^{\dagger}_{R/L}(x_\alpha,t')\}\rangle$ we rewrite the particle current as
\begin{eqnarray}\label{mix}
 I(t)/e=\Re[\sum_{j=R,L;\sigma}t_{j\sigma\alpha}e^{-i\mu t}G_{\sigma,j,\alpha}^{<}(t,t)].
\end{eqnarray}

 This lesser mixed Green's function $G_{\sigma,j,\alpha}^{<}(t,t)$ is obtained by perturbation theory as
\begin{eqnarray}\nonumber
&&G_{\sigma,R/L,\alpha}(t,t')=\sum_{l=0}^{\infty}\frac{(-i)^{l+1}}{l!}\int_c d\tau_1\ldots\int_c d\tau_l\langle T_c\{\\\label{dg}&&\gamma_{\sigma}(y_\alpha,t)H_{int}(\tau_1) \ldots
H_{int}(\tau_l)\psi^{\dagger}_{R/L}(x_\alpha,t')\}\rangle
\end{eqnarray}
In applying the Wick theorem in the Eq.(\ref{dg}) we should also include all possible four fermions interactions term ($u_2$ and $u_4$ term in the edge states Hamiltonian) between any two fermions operators. We use the spinless bosonization\cite{TG} as a way to sum up all orders of perturbations in the four fermions interactions on the Keldysh contour. The edge state correlators evaluated this way is thus fully dressed by the four fermions interactions in our treatment and we do not specify this aspect in the expression of Eq.~(\ref{dg}). 

We bosonize the helical Luttinger liquids lead operators by writing the fermion fields as:
\begin{eqnarray}\nonumber
&&\psi_{R}(x)=\frac{1}{\sqrt{2\pi a_0}}\eta_{R}e^{-i\sqrt{4\pi}\phi_R(x)},\\\label{boson}
&&\psi_{L}(x)=\frac{1}{\sqrt{2\pi a_0}}\eta_{L}e^{i\sqrt{4\pi}\phi_L(x)},
\end{eqnarray} 
with $\eta_{R/L}$ as the Klein factor chosen to satisfy the fermion anti-commutation rule and $a_0$ as the lattice spacing cutoff for the linear spectrum. We define the bosonic fields $\Phi,\Theta=\phi_{L}\pm\phi_{R}$ and rewrite $H_0=H_{L}+H_{M_0}$ and $H_{T}$ as
\begin{eqnarray}\nonumber
H_0&=&\frac{v}{2}\int_{-\infty}^{\infty} dx :[K(\partial_x \Theta)^2+\frac{1}{K}(\partial_x \Phi)^2]:\\\nonumber
&+&i \sum_{\sigma}v_{M,\sigma}\int_0^L dx \gamma_{\sigma}(x)\partial_x \gamma_{\sigma}(x)\\\nonumber
H_T&=&\sum_{\sigma\alpha}i\gamma_{\sigma}\Big(e^{-i\mu t}\big(\frac{t_{R\sigma\alpha}}{\sqrt{2}}e^{i\sqrt{4\pi}\phi_{R}(x_\alpha)}\eta_{R}^{\dagger}\xi_{R\sigma\alpha}^{\ast}\\\label{Ham}&+&\frac{t_{L\sigma\alpha}}{\sqrt{2}}e^{-i\sqrt{4\pi}\phi_{L}(x_\alpha)}\eta_{L}^{\dagger}\xi_{L\sigma\alpha}^{\ast}\big)+h.c.\Big)
\end{eqnarray}
with Luttinger parameter $K=\sqrt{\frac{2 \pi v_F+u_4-u_2}{2 \pi v_F+u_4+u_2}}$ and velocity $v=v_F\sqrt{(1+\frac{u_4}{2\pi v_F})^2-(\frac{u_2}{2\pi v_F})^2}$.
Eq.~(\ref{Ham}) serves as the main Hamiltonian for computing the tunneling current in section \ref{current}. For a single tunneling point contact with time reversal symmetry preserved we set $t_{R\uparrow\alpha}=t_{L\downarrow\alpha}=\bar{t}_\alpha$ and otherwise zero. We discuss why other relevant perturbations $\delta H$ are not important in this time reversal preserved system in the next section.

\section{Scaling analysis}
Following the discussions in Ref.~\onlinecite{Fidkowski} for a single tunneling junction located at $x=0$, the most relevant terms $\delta H$ other than the tunneling term $H_T$ are
\begin{eqnarray}
\delta H&=&V_1[\psi_R^{\dagger}(0)\psi_R(0)+\psi_L^{\dagger}(0)\psi_L(0)]\\\nonumber
&+&[V_2\psi_R^{\dagger}(0)\psi_L(0)+\Delta\psi_R(0)\psi_L(0)+\text{h.c.}].
\end{eqnarray}
Here, the $V_1$ terms represent the chemical potential change due to the presence of the tunneling junction (also called quantum point contact). The $V_2$ terms stand for backscattering due to the point contact and $\Delta$ is the Cooper pair gap magnitude induced at $x=0$ via proximity effect.\cite{Fidkowski,YW} Rewriting the fermionic operators via Eq.~(\ref{boson}) we get $\delta H$ in bosonized form as
\begin{eqnarray}\nonumber
\delta H&=&\frac{V_1}{\sqrt{\pi}}\partial_x\Phi(0)-\frac{V_2}{\pi a_0}\sin\left(\sqrt{4\pi}\Phi(0)\right)\\
&+&\frac{|\Delta|}{\pi a_0}\sin\left(\sqrt{4\pi}\Theta(0)-\phi\right),
\end{eqnarray}
with $\Delta=|\Delta|e^{i\phi}$. The $V_1$ terms can be absorbed in the definition of $\Phi(x)$ by the shift $\Phi(x)\rightarrow \Phi(x)-\frac{KV_1}{2v\sqrt{\pi}}\text{sgn}(x)$. For the rest of the terms in $H_T$ and $\delta H$ the scaling dimensions around $H_0$ are\cite{YW}
\begin{eqnarray}\nonumber
&&D[\bar{t}]=\frac{1}{4}\left(K+\frac{1}{K}\right)+\frac{1}{2},\\\nonumber
&&D[V_2]=K,\\\label{rga}
&&D[\Delta]=1/K.
\end{eqnarray}
The term $\frac{1}{2}$ in $D[\bar{t}]$ comes from the scaling dimension of helical Majorana modes $D[\gamma]=\frac{1}{2}$, which is the same as chiral ones, assuming its spectrum is continuous (or the boundary of the topological superconductor being infinite). For a time reversal symmetric Hamiltonian the backscattering term proportional to $V_2$, being the 
only relevant term in the repulsive interaction regime ($0<K\le 1$), is forbidden. For repulsive interactions $1\le D[\bar{t}]\le D[\Delta]$ and thus the 
most important terms (marginally relevant) in perturbation for $H_T+\delta H$ around $H_0$ is the tunneling term $H_T$. 

For the short topological superconductors considered in this paper, the helical Majorana edge states become discretized and $D[\gamma]\simeq 0$. Under this approximation $D[\bar{t}]\simeq \frac{1}{4}\left(K+\frac{1}{K}\right)$ becomes relevant for $2-\sqrt{3}<K<1$, while $D[\Delta]$ stays irrelevant in the repulsive regime, indicating the same quantum phase transition (metallic to insulating) as for helical Luttinger liquids connected via a quantum dot\cite{sp} in the repulsive regime.

For different geometries, such as a Luttinger liquid terminated at a Majorana zero mode end state\cite{Fidkowski} or helical Luttinger liquid connected to a time-reversal breaking topological superconductor (with chiral Majorana modes as its edge state),\cite{YW} the backscattering $V_2$ term is relevant for $K<1$ and the low-energy physics is determined by a new fixed point Hamiltonian:\cite{YW}
\begin{eqnarray}
H_0^{'}=H_0-\frac{V_2}{\pi a_0}\sin\left(\sqrt{4\pi}\Phi(0)\right),
\end{eqnarray} 
which fixes the value of $\Phi(0)=\sqrt{\pi}/4$ for $V_2>0$. Under this constraint the scaling dimension of tunneling term $D[\bar{t}]$ with $D[\gamma]\simeq 0$ becomes $D[\bar{t}]\simeq \frac{1}{2K}$ and is relevant for $1/2<K<1$, giving rise to the transition between perfect normal and perfect Andreev reflection  at $K=1/2$ in this system.\cite{Fidkowski} The transition from perfect normal to perfect Andreev reflection is shown as insulating to metallic transition in the charge transport. The key difference from our setup is the different scaling behavior (different power law dependence), controlled by the density-density interaction strength in the helical Luttinger liquid, in the differential conductance as a function of bias voltage or temperature.

\section{Evaluating the current}\label{current}
In this section we carry out the calculation of the current for a single point contact and a double point contact at zero temperature. We start by finding analytic expressions for the helical Luttinger liquid and dressed helical Majorana modes Green's functions. From there we compute the current numerically by using Eqs.~(\ref{mix}) and (\ref{dg}), 
and thus obtain the current-voltage relation numerically. By taking the derivative numerically we get the differential conductance as a function of voltage. We find a metallic to insulating quantum phase transition (near zero bias) with increasing repulsive interaction, and less universal patterns owing to the interference nature in the case of double point contacts. 

As a side remark, notice that the computation carried out here is not the one loop RG calculations mentioned in the previous section. We perform a diagramatic based resummation of perturbative terms and the evaluated differential conductivities depend explicitly on the choice of linear momentum cutoff $\Lambda$. The choice of $\Lambda$ depends on the particular realizations of the helical Luttinger modes, i.e. material dependent. The cutoff dependence, as shown in Appendix \ref{A0}, is consistent with the trend we expect from usual higher order (two loops or more) of RG calculations. That is, for larger cutoff $\Lambda$, the deviations from what we expect from lowest RG analysis are larger. In the rest of the paper we choose $\Lambda=10^{-2}\epsilon_F$ as a typical value of modeling the linearization of some quadratic bands at the Fermi surface, or the band touching point where the edge states of 2DTI become mixed with the bulk band.   

\subsection{Single point contact}
We start with single point contact between the helical Luttinger liquid and helical Majorana modes realized in a time-reversal symmetric topological superconductor. 
From Ref.~\onlinecite{sp} the Keldysh component of bare (uncoupled) lead Green functions, defined as $G_{\psi_{L/R}}(\tau,\tau')=-i\langle T_c\{\psi_{L/R}(\tau)\psi_{L/R}^{\dagger}(\tau')e^{-i\mu_{L/R}(\tau'-\tau)}\}\rangle$, expressed in frequency space at zero temperature are\cite{footnote2}
\begin{eqnarray}\label{gsingle}
G_{\psi_{L/R}}^{++}(\omega)&=&\frac{a_0^{2\kappa}}{4\pi^2 v^{2\kappa}}\frac{\Gamma(\kappa)^2}{\Gamma(2\kappa)}|\omega-\mu|^{2\kappa-1}\\\nonumber
&\times&\left(\tilde{h}(\kappa)\theta(\omega-\mu)-\tilde{h}(\kappa)\theta(\mu-\omega)\right)\\\nonumber
G_{\psi_{L/R}}^{--}(\omega)&=&\frac{a_0^{2\kappa}}{4\pi^2 v^{2\kappa}}\frac{\Gamma(\kappa)^2}{\Gamma(2\kappa)}|\omega-\mu|^{2\kappa-1}\\\nonumber
&\times&\left(-\tilde{h}^{\ast}(\kappa)\theta(\omega-\mu)+\tilde{h}^{\ast}(\kappa)\theta(\mu-\omega)\right)\\\nonumber
G_{\psi_{L/R}}^{+-}(\omega)&=&\frac{a_0^{2\kappa}}{v^{2\kappa}}\frac{i}{\Gamma(2\kappa)}|\omega-\mu|^{2\kappa-1}\theta(\mu-\omega)\\\nonumber
G_{\psi_{L/R}}^{-+}(\omega)&=&\frac{a_0^{2\kappa}}{v^{2\kappa}}\frac{-i}{\Gamma(2\kappa)}|\omega-\mu|^{2\kappa-1}\theta(\omega-\mu)
\end{eqnarray}
Here $\kappa=\frac{1}{4}(K+1/K)$ and $\tilde{h}(\kappa)=2e^{-\pi i \kappa}\sin(\pi\kappa)\Gamma(1-\kappa)^2$ and plus/minus sign on $G_{\psi_{L/R}}$ indicates its labeling on the Keldysh contour (with $G_{\psi_{L/R}}^{++}$ as time ordered and $G_{\psi_{L/R}}^{--}$ as anti-time ordered). By relabeling the spin index in $\gamma_\sigma$ by the left/right-movers label of the Luttinger lead operator, the steady state charge current is expressed as
\begin{eqnarray}\nonumber
\langle\hat{I}\rangle&=&e\Re\Big[\sum_{n,m;j=L,R}\frac{t_{j,n}t_{j,m}^{\ast}}{2}\int d\omega \big(G_{\gamma_{j,nm}}^R(\omega)G_{\psi_j}^{<}(\omega)\\ \label{hcurr}
&+&G_{\gamma_{j,nm}}^{<}(\omega)G_{\psi_j}^{A}(\omega)\big)\Big].
\end{eqnarray}

Here $n$, $m$ denote the discrete energy levels in the finite size helical Majorana modes. Eq.(\ref{hcurr}) follows from maintaining the structure of first order expansion in Eq.(\ref{dg}) and resum all higher order terms through the "dressed" helical Majorana Green's function.
The retarded helical Majorana Green's function contains higher order terms through inclusion of self energy terms:
\begin{eqnarray}\label{eq14}
G_{\gamma_{j,nm}}^{R}(\omega)&=&G_{\gamma_{j,nm}}^{(0)R}(\omega)\\\nonumber&+&\sum_{l,l'}G_{\gamma_{j,nl}}^{(0)R}(\omega)\Sigma_{j,ll'}^R(\omega)G_{\gamma_{j,l'm}}^{R}(\omega). 
\end{eqnarray}
Here the "bare" retarded helical Majorana Green's function is
$G_{\gamma_{j,nm}}^{(0)R}(\omega)=\delta_{n,m}/(w-\epsilon_{n,j}+i0^+)$ with $\epsilon_{n,j}=\hbar v_{M}\text{sgn}(j) \frac{2\pi n}{L}$ (with $\text{sgn}(j)=+/-$ for $L/R$), and the retarded self energy is $\Sigma_{j,nm}^R(\omega)\equiv\frac{t_{j,n}t_{j,m}^{\ast}}{2}G_{\psi_j}^R(\omega)$ given by the Dyson equation. The dressed helical Majorana lesser Green's function is
$G_{\gamma_{j,nm}}^{<}(\omega)=G_{\gamma_{j,nl}}^{R}(\omega)\Sigma_{j,ll'}^{<}(\omega)G_{\gamma_{j,l'm}}^{A}(\omega)$ with $\Sigma_{j,nm}^{<}(\omega)\equiv\frac{t_{j,n}t_{j,m}^{\ast}}{2}G_{\psi_j}^{<}(\omega)$. Similar expressions hold for $G_{\gamma_{j,nm}}^{A}(\omega)$ and
$G_{\gamma_{j,nm}}^{>}(\omega)$.

\begin{figure}
\includegraphics[width=.9\columnwidth]{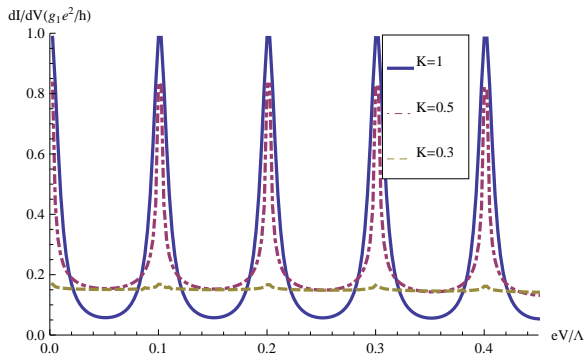}
\caption{Differential conductance as a function of voltage for different Luttinger parameters in the helical Luttinger lead connected by single quantum point contact with
grounded helical Majorana modes. The Luttinger parameters are: $K=1$ (blue solid), $K=0.5$ (purple dot dashed), $K=0.3$ (brown dashed). Other parameters are $\bar{t}=0.05\Lambda$, the length of the edge of topological superconductor $L=10^3a_0$, and linear spectrum cutoff $\Lambda=10^{-2}\epsilon_F=10\hbar v_M\frac{2\pi}{L}$.}
 \label{f3}
\end{figure}

In Eq.~(\ref{hcurr}) the summation over integers $n$, $m$ (and $l$,$l'$ in Eq.(\ref{eq14})) refers to the sum over discrete Majorana modes energy level indices.
From this the expression for the current is related to the evaluation of $G_{\psi_j}^{<}(\omega)=G_{\psi_j}^{+-}(\omega)$, $G_{\psi_j}^{A}(\omega)=G_{\psi_j}^{++}(\omega)-G_{\psi_j}^{-+}(\omega)$, and the aforementioned dressed helical Majorana Green's functions. In this paper we assume the energy difference between different Majorana modes is sufficient large (greater than the broadening effect coming from coupling with the Luttinger lead) such that $\Sigma_{j,nm}(\omega)\simeq\Sigma_{j}(\omega)\delta_{n,m}$ to simplify the calculation. In other words, we consider the helical topological insulator as short, such that the finite size makes the energy difference between discrete Majorana modes sufficient large so that the overlap between them is negligible. Under this assumption the helical Majorana Green's functions $G_{\gamma_{j,nm}}^{>}(\omega)$ and $G_{\gamma_{j,nm}}^{R}(\omega)$ are diagonal and the analytic expression for Eq.~(\ref{hcurr}) is obtained. From there we evaluate the current numerically and obtain its relation with bias voltage $V=(\mu-0)/e$ to evaluate the differential conductance. The results for single tunneling junction, with different Luttinger parameters $K$ indicating different interaction strengths, are shown in Fig.~\ref{f3}. 

In Fig.~\ref{f3} we see the perfect transmission (maximum differential conductance) at zero voltage for a noninteracting ($K=1$) helical Luttinger liquid. It originates from the perfect Andreev reflection between the metal and superconductor mediated by the helical Majorana modes.\cite{Law} The Majorana modes inside the superconducting gap serve as resonance levels which facilitates the Andreev reflection process and give a differential conductance value $g_1e^2/h$ with $g_1=2\times 2 = 4$, reflecting particle-hole and spin symmetry. This perfect transmission signature is used to identify the Majorana zero modes in the nanowire experiments.\cite{Mourik} The periodic peaks at finite bias voltages, similar to the case of tunneling measurement chiral Majorana modes discussed in Ref.~[\onlinecite{Law}], come from discrete Majorana energy levels with energy difference (peak intervals) set by the physical size of the edge of helical topological superconductor. For (repulsive) interacting lead the general feature is the suppression of the resonant conductance peaks, both in peak magnitude and width, and the spreading out of spectral weight away from the resonance levels.\cite{sp} 

\begin{figure}
\includegraphics[width=.9\columnwidth]{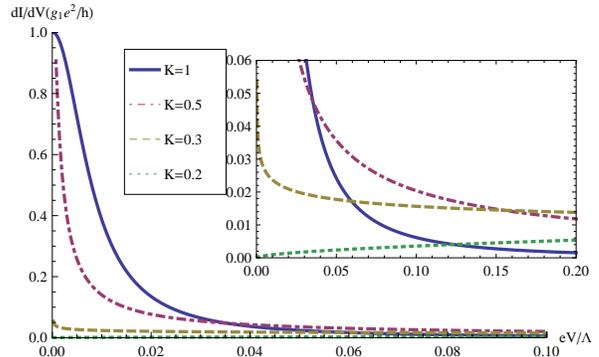}
\caption{Differential conductance as a function of voltage for different Luttinger parameters with a single Majorana level: $K=1$ (blue solid), $K=0.5$ (purple dot dashed), $K=0.3$ (brown dashed),$K=0.2$ (green dotted). Other parameters are the same as in Fig.~\ref{f3}.}
 \label{f4}
\end{figure}

The spreading of spectral weight makes the transition from perfect Andreev reflection to perfect normal reflection more difficult to observe at smaller Luttinger parameter $K$. This is because the Majorana levels begins to merge together (as shown for $K=0.3$ case in the Fig.~\ref{f3}), violating our starting assumption that the levels are sufficient far apart. To illustrate this kind of metallic (perfect Andreev reflection) to insulating (perfect normal reflection) behavior in this single tunneling junction context, we plot the differential conductance for a single level (zero energy) helical Majorana mode in the Fig.~\ref{f4}. We see that the transition takes place between $K=0.3$ and $K=0.2$ with marked tendency differences between the two at finite bias. 

The scaling analysis mentioned in the previous section for single Majorana mode gives $D[\bar{t}]=(K+1/K)/4$, resulting in a critical Luttinger parameter $K_{cr}=2-\sqrt{3}$ for the repulsive helical Luttinger lead. This scaling/criticality behavior is the same as the case for two helical Luttinger liquids connected (with particle-hole symmetry imposed, or $\mu_1=-\mu_2=eV/2$) by a noninteracting single level quantum dot discussed in Ref.~[\onlinecite{sp}]. The vanishing charge transport at zero bias below $K=K_{cr}$ corresponds to the transition point where some quantum information stored by qubits formed by Majorana modes
is maintained and does not decohere completely.\cite{shHo} In general, for charge transport we can formally make analogy between the two helical Luttinger leads with particle-hole symmetric driven voltage connected via a noninteracting multi-level quantum dot system with our single helical Luttinger lead connected with the helical Majorana modes.

Note that once introducing time-reversal-invariance breaking terms\cite{YW} or different ways\cite{Fidkowski} of connecting the Majorana modes, the critical behavior in the charge transport could occur at different $K_{cr}$. For example, if we replaced the helical Majorana modes by the chiral Majorana modes (the edge state of a topological superconductor with broken time reversal symmetry),\cite{YW} similar metallic to insulating behavior is seen but with scaling behavior controlled by $D[\bar{t}]=1/2K$, or $K_{cr}=1/2$. Thus for single Majorana modes (or other modes sufficiently far apart such that the overlap is not significant) different scaling behaviors in the single tunnel junction transport reveals a great deal of information about the boundary conditions imposed on the Luttinger liquid.
  
In a real experiment there is a finite length region where tunneling between the Majorana modes and helical lead occurs. Assuming a separable form for the spatial dependence of tunneling term\cite{Martin,Sassetti} the analytic expressions for weak tunneling current are obtained for the case of leads made of same type of material. For tunneling between helical Luttinger leads, the power-law dependence is modified for an extended contact compared to the case of a point like contact.\cite{Sassetti} Extension of this formulation to infinite-size helical Majorana lead case seems to be straightforward
but not so easy for the case of a finite helical Majorana lead. Therefore, we proceed with the simpler case: the case of two quantum point contacts.     

\subsection{Double point contact}
Proposals for double point contact setups are mainly related to the study of quantum interference effects\cite{Wen,Rosenow} and the quasi-particles statistics of the edge states.\cite{Stern,Slingerland,Slingerland2} Similar setups for the helical edge states have been discussed\cite{Recher,Rizzo,Huang,Orth13} with the possible applications for electronic means of spin pumping\cite{Dolcini,Natan}. Different types of interferometers have also been proposed for heterostructures of topological superconductor and normal metal/edge states of topological insulators such as the Majorana Dirac converter.\cite{Liu,Beri,Beenakker} Here, the double point contact between the helical Luttinger liquid and the Majorana modes is yet another type of heterostructure showing the quantum interference which is analogous to the two point source interference in optics.

For two point contacts the total current passing through those contacts is
$\langle\hat{I}\rangle=\langle\hat{I}_{x_1}+\hat{I}_{x_2}\rangle$. Without loss of generality we choose $x_1=0$, $x_2=x$, $y_1=0$ and $y_2=y$. Here $x_i$ denotes spatial coordinate of the fermion operators in the helical Luttinger lead and $y_i$ denotes that of the helical Majorana operators in the topological superconductor. We evaluate the current $\langle\hat{I}\rangle$ via perturbations on the tunneling term $H_T$ on the Keldysh contour. To simplify the notation we
denote $G_{(\psi/\gamma)_\alpha}^{(0)}$ as the bare (unperturbed) helical fermion/Majorana mode with chiral (or spin) and position index $\alpha$ and concentrate on the structure of perturbation in the Dyson equation without bookkeeping the Keldysh contour labels on the Green's functions for the moment. We get:
\begin{widetext}
\begin{eqnarray}\label{eq15}
&&G_{\gamma_\alpha}=G_{\gamma_\alpha}^{(0)}+G_{\gamma_\alpha}^{(0)}|t_\alpha|^2 G_{\psi_\alpha}G_{\gamma_\alpha}+G_{\gamma_\alpha}^{(0)}t_\alpha G_{\psi_\alpha\psi_\beta}t_\beta^{\ast}G_{\gamma_\beta\gamma_\alpha}
+G_{\gamma_\alpha\gamma_\beta}^{(0)}t_\beta G_{\psi_\beta\psi_\alpha}t_\alpha^\ast G_{\gamma_\alpha}+G_{\gamma_\alpha\gamma_\beta}^{(0)}|t_\beta|^2 G_{\psi_\beta}G_{\gamma_\beta\gamma_\alpha}\\\label{eq16}
&&G_{\gamma_\alpha\gamma_\beta}=G_{\gamma_\alpha\gamma_\beta}^{(0)}+G_{\gamma_\alpha}^{(0)}t_\alpha G_{\psi_\alpha\psi_\beta}t_\beta^{\ast}G_{\gamma_\beta}+G_{\gamma_\alpha\gamma_\beta}^{(0)}|t_\beta|^2 G_{\psi_\beta}G_{\gamma_\beta}
+G_{\gamma_\alpha\gamma_\beta}^{(0)}t_\beta G_{\psi_\beta\psi_\alpha}t_\alpha^\ast G_{\gamma_\alpha\gamma_\beta}+G_{\gamma_\alpha}^{(0)}|t_\alpha|^2 G_{\psi_\alpha}G_{\gamma_\alpha\gamma_\beta}
\end{eqnarray}
\end{widetext}

Note that we do not have spin flip process in the tunneling term, and the $G_{\psi_\beta\psi_\alpha}$ or $G_{\gamma_\beta\gamma_\alpha}$ are diagonal in spin space and functions of differences in the spatial coordinate. In these simplified notations the current at position $x_\alpha$ of the helical Luttinger lead coordinate is 
\begin{eqnarray}\label{gcur}
\langle\hat{I}_{x_\alpha}\rangle=e\Re[\int d\omega\left(|t_\alpha|^2 G_{\gamma_\alpha}G_{\psi_\alpha}+t_\alpha t_{\beta}^{\ast}G_{\gamma_\alpha \gamma_\beta}
G_{\psi_\beta \psi_\alpha}\right)]
\end{eqnarray}
It is easy to check that above formula gives the single point contact result (\ref{hcurr}) by taking $t_\alpha=\bar{t}/\sqrt{2}$, $t_\beta=0$, and with the Langreth theorem\cite{Langreth} (to denote the contour order). Following the same recipes the current for two point contacts with $t_\alpha=\bar{t}_1/\sqrt{2}$ and $t_\beta=\bar{t}_2/\sqrt{2}$ is then expressed as:
\begin{widetext}
\begin{eqnarray}\label{eq18}
&&\langle\hat{I}\rangle=\sum_{j=\pm 1}\left(\langle\hat{I}_{x_1,j}\rangle + \langle\hat{I}_{x_2,j}\right)\rangle\\\nonumber
&&\langle\hat{I}_{x_1,j}\rangle=\frac{e}{2}\Re[\int d\omega\Big(|\bar{t}_1|^2 (G_{\gamma_j}^R(\omega)G_{\psi_j}^{<}(\omega)+G_{\gamma_j}^{<}(\omega)G_{\psi_j}^{A}(\omega)) 
+ \bar{t}_1 \bar{t}_2^{\ast}(G_{\gamma_j}^R(\omega,-y_{12})G_{\psi_j}^{<}(\omega,x_{12})+G_{\gamma_j}^{<}(\omega,-y_{12})G_{\psi_j}^{A}(\omega,x_{12}))\Big)]\\\nonumber
&&\langle\hat{I}_{x_2,j}\rangle=\frac{e}{2}\Re[\int d\omega\Big(|\bar{t}_2|^2 (G_{\gamma_j}^R(\omega)G_{\psi_j}^{<}(\omega)+G_{\gamma_j}^{<}(\omega)G_{\psi_j}^{A}(\omega)) 
+ \bar{t}_2 \bar{t}_1^{\ast}(G_{\gamma_j}^R(\omega,y_{12})G_{\psi_j}^{<}(\omega,-x_{12})+G_{\gamma_j}^{<}(\omega,y_{12})G_{\psi_j}^{A}(\omega,-x_{12}))\Big)]
\end{eqnarray}
\end{widetext}

\begin{figure}
\includegraphics[width=1\columnwidth]{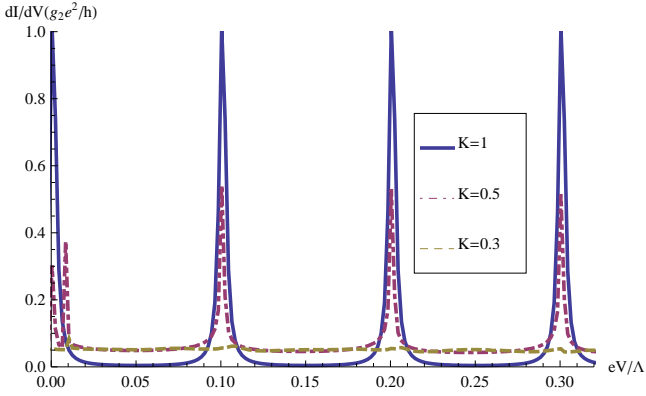}
\caption{Differential conductance v.s. voltage for different Luttinger parameters $K=1$, $K=0.5$, and $K=0.3$ with separation distance $x_{12}=y_{12}=10a_0$. We choose the tunneling term $\bar{t}_1/\Lambda=\bar{t}_2/\Lambda=0.01$, and the length of the helical Majorana modes $L=10^3a_0$ with $a_0$ denoting lattice spacing. $\Lambda=10^{-2}\epsilon_F=10\hbar v_M\frac{2\pi}{L}$ is the linear spectrum cutoff in the helical Luttinger lead.}
 \label{f5}
\end{figure}
\begin{figure}
\includegraphics[width=1\columnwidth]{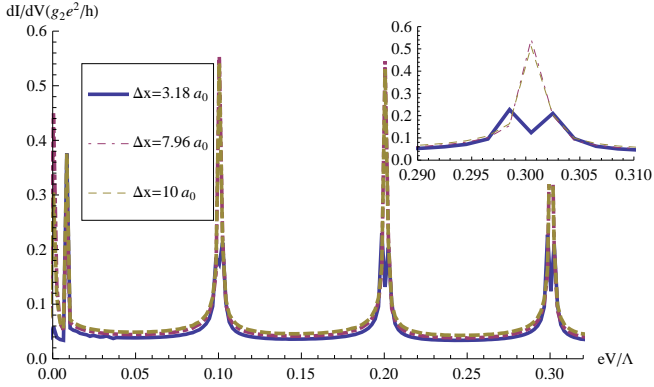}
\caption{Differential conductance v.s. voltage for Luttinger parameter $K=0.5$ with different separation lengths. We choose $x_{12}=y_{12}$ with $x_{12}=3.18a_0$ (blue solid), $x_{12}=7.96a_0$ (purple dot dashed), and $x_{12}=10a_0$ (brown dashed). Other parameters: $\bar{t}_1/\Lambda=\bar{t}_2/\Lambda=0.01$, $L=10^3a_0$, and $\Lambda=10^{-2}\epsilon_F=10\hbar v_M\frac{2\pi}{L}$. Top right inset shows the enlarged figure for peaks around $eV/\Lambda=0.3$.}
 \label{f6}
\end{figure}

\begin{figure}
\includegraphics[width=1\columnwidth]{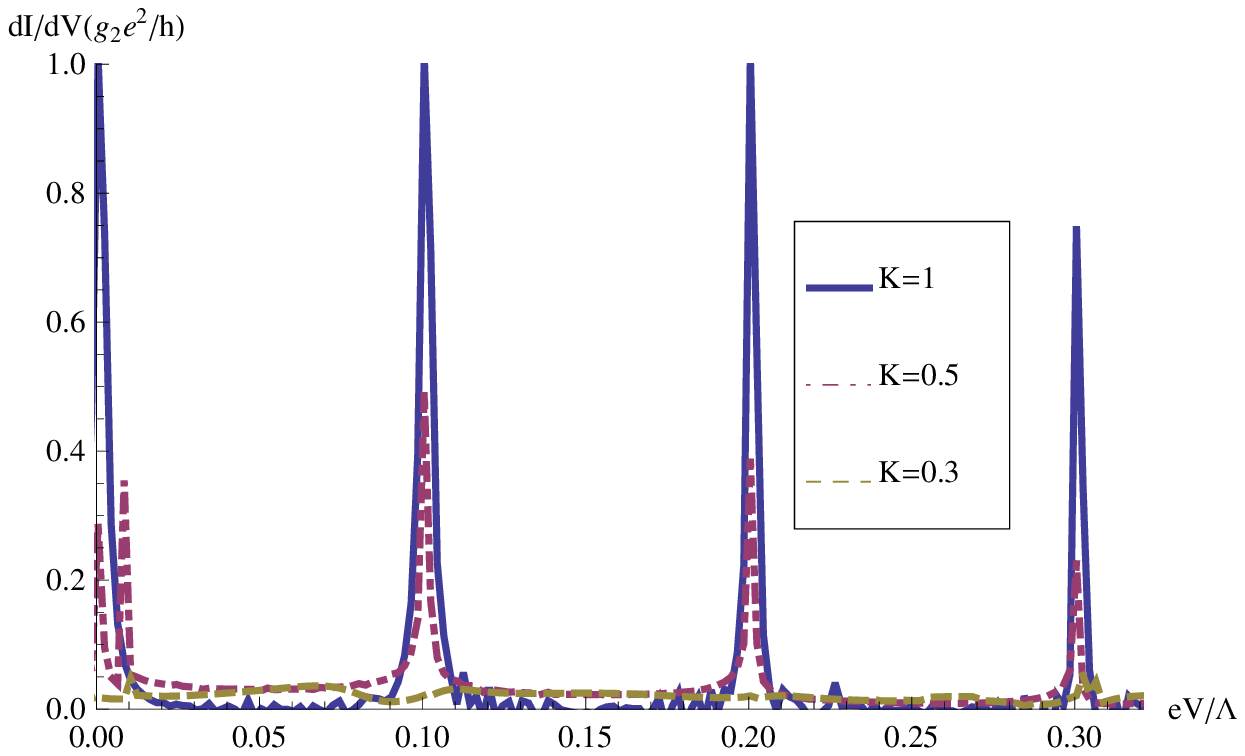}
\caption{Differential conductance v.s. voltage for different Luttinger parameters $K=1$ (blue solid), $K=0.5$ (purple dot dashed), and $K=0.3$ (brown dashed) with separation distance $x_{12}=y_{12}=10^2a_0$. We choose the tunneling term $t_1/\Lambda=t_2/\Lambda=0.01$, and the length of the helical Majorana modes $L=10^3a_0$ with $a_0$ denoting lattice spacing. $\Lambda=10^{-2}\epsilon_F=10\hbar v_M\frac{2\pi}{L}$ is the linear spectrum cutoff in the helical Luttinger lead.} \label{f7}
\end{figure}

\begin{figure}
\includegraphics[width=1\columnwidth]{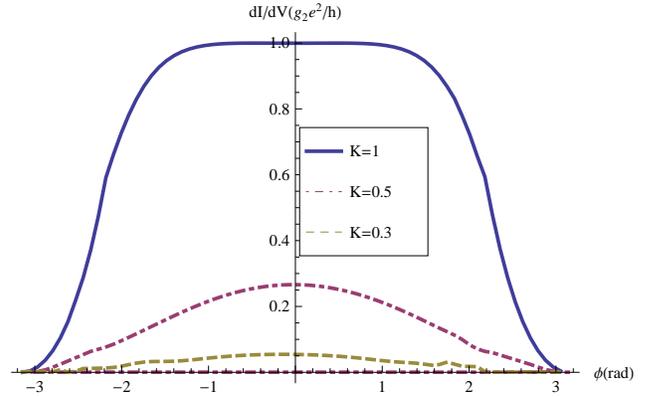}
\caption{Differential conductance v.s. tunneling amplitude phase difference $\phi$ for different Luttinger parameters $K=1$ (blue solid), $K=0.5$ (purple dot dashed), and $K=0.3$ (brown dashed). $t_1=t_2 e^{i\phi}$ with $|t_1|/\Lambda=0.01$. Other parameters: $x_{12}=y_{12}=10a_0$, $\Lambda=10^{-2}\epsilon_F=10\hbar v_M\frac{2\pi}{L}$, and $L=10^3a_0$.}
 \label{f8}
\end{figure}

Here $G_{\gamma_j}^{(0)R}(\omega,y)=\sum_{n}\frac{e^{-\text{sgn}(j)i \frac{2\pi n}{L}y}}{\omega-\epsilon_{n,j}+i\eta}$ and $G_{\gamma_j}^{(0)<}(\omega,y)=2\pi i\sum_{n}e^{-i \text{sgn}(j)\frac{2\pi n}{L}y}\theta(-\omega)\delta(\epsilon_{n,j}-\omega)$ are the unperturbed retarded and lesser Green's function for helical Majorana modes, label $j=\pm 1$ denotes left/right moving mode, and $x_{12}=x_1-x_2$ and $y_{12}=y_1-y_2$ are the spatial coordinate differences. Following Eqs.~(\ref{eq15}) and (\ref{eq16}) and the Langreth rule we obtain the various dressed Majorana Green's functions and unperturbed helical fermions Green's functions needed for evaluating the current. The derivations and analytic expressions for various Green's functions are shown in the Appendix \ref{AA} and \ref{AB}. With the analytic expressions shown in the Appendices, we perform numerical integrals to compute the current (\ref{eq18}) and obtain the differential conductance by taking numerical derivatives with respect to the source drain voltage $V$. The results are shown in the Figs.~\ref{f5}--\ref{f8}.

We chose a small separation length ($x_{12}=y_{12}=10^{-2}L$) between the two contacts in the Fig.~\ref{f5} and fixed the tunneling strengths of the two point contacts to be identical. The interference effect due to two point contacts for the weakly interacting lead ($K\simeq 1$) is not apparent, and the resonance structure is similar to the single point contact. For a noninteracting lead ($K=1$ or blue solid line in Fig.~\ref{f5}) the differential conductance reaches its maximum value $g_2 e^2/h$ with $g_2=2\text{(particle hole)}\times 2\text{(spin)}\times 2\text{(2 tunneling points)}=8$ when the chemical potential of the helical lead is in line with the discrete Majorana energy levels. For a helical lead with stronger repulsion (say $K=0.5$ or purple dashed line in Fig.~\ref{f5}) we see features similar to the single point contact (with shrinking peak width and height at resonance value and transfer of spectral weight away from 
the resonance) and the effect of interference between two point contacts. Around zero bias the peak splits into two, similar to the physics of Fano resonance, and slight modulations in the resonance positions in other finite voltage peaks. To further study the interference effect we fix $K=0.5$ and plot different separation lengths (still keeping $x_{12}=y_{12}$ and $x_{12}/L \sim 10^{-2}$) in Fig.~\ref{f6}. We see the subpeak structure (see the inset of Fig.~\ref{f6}) also emerges nearby finite voltage resonance peaks with peak heights at a fixed voltage depending on the separation distance. This kind of subpeak structure mainly comes from the change in the real part of self energy correction on the Majorana Green's function, which emerges with the cancellation of fast oscillating term related to $e^{ik_F sgn[j] x_{12}}$ in the $G_{\psi_{\alpha}\psi_{\beta}}$ in the Appendix \ref{AB} from different orientations. 

For longer separation distance the interference effect also brings a change in the peak heights. To demonstrate this we choose $x_{12}=y_{12}=0.1L$ in Fig.~\ref{f7} such that the separation distance is one tenth of the linear dimension of the Majorana modes. Other than the subpeak structure seen for $K=0.5$ case we now also see modulations in the resonance peak heights. This larger envelope (modulation with large voltage range) is associated with the separation length scale being comparable with the Majorana system size. For off resonance region ($K=1$ plot with voltage between $eV/\Lambda=0$ to $0.1$ in Fig.~\ref{f7}, for example) we also see small oscillations around zero which is attributed to the inaccuracy of numerical integrals for fast oscillating functions.

For general double point contacts we could have different tunneling amplitudes $t_1$, $t_2$ and different separation distances $|x_{12}|\neq |y_{12}|$. For $|t_1|\gg|t_2|$ or $|t_1|\ll|t_2|$ the transmitted current is dominated by one of the point contact, and the result is basically the same as that of the single point contact. For $|x_{12}|\neq |y_{12}|$ but with $|t_1|=|t_2|$ and $|x_{12}|\approx |y_{12}|$ the general features are similar to what we have mentioned in this section. Here we discuss the case of identical separation length $|x_{12}|=|y_{12}|=10a_0$ but with different tunneling amplitude $t_1=t_2 e^{i\phi}$. We plot differential conductance around zero bias as a function of the tunneling phase difference $\phi$ for different Luttinger parameters in Fig.~\ref{f8}. For small separation distance chosen here $\phi=\pi$ or $t_1=-t_2$ leads to almost complete cancellation of the resonance peak. For larger separation distance the general feature is the same (decreasing $dI/dV$ with increasing $\phi$) but with finite conductance even at $\phi=\pi$.

\section{Conclusion}
We have investigated the charge transport between a helical Luttinger liquid and a system of helical Majorana fermions coupled by single and double quantum point contacts. The helical Luttinger liquid is realized as the one dimensional edge state of a thin film of a 3D topological insulator with the inclusion of short-range repulsion. The helical Majorana fermion could be realized in noncentrosymmetric topological superconductor or proximity-induced effective topological superconductor with time reversal symmetry. For a single tunneling point contact we find that perfect Andreev reflection occurs only for a noninteracting helical lead. Increasing the repulsive interaction strength leads to the suppression of the differential conductance on resonance and shifts the weight away from resonance. This feature is similar to the case of two Luttinger leads connected by a noninteracting quantum dot\cite{sp} with particle hole-symmetric bias voltage ($\mu_1=-\mu_2=eV/2$).

We then studied the case of two quantum point contacts. For small separation distance ($x_{12}\ll L$ with $L$ being the size of the edge of the topological superconductor), the interference from the two point contacts strongly changes the shape of the individual resonance peaks but does not affect the overall magnitude at different Majorana mode energies. At larger separation distance ($x_{12} \sim 10^{-1}L$) we observe modulations in the magnitude and shape of individual resonance peaks resulting from two point interference.
 
In a real experimental setup, the point contact may not be perfect, in which case an extended contact may provide a better description.\cite{Martin,Sassetti} The analytic results of the perturbation theory in the tunneling get more complicated with an increased number of tunneling channels, as shown for the case of two point contacts in this paper. We conjecture, based on our result at small separation distance, that with sufficiently small size of this extended point contact ($\Delta x\sim \Delta y \ll L$), the overall transport behavior will be similar to the single point contact. The detailed scaling behavior\cite{Sassetti} or the shape of the individual resonance peak can be different and the transport signature gets modified by the interaction more significantly. This can also be viewed as a generalization of the scaling behavior change due to the modification of the boundary conditions as mentioned in Ref.~[\onlinecite{YW}]. 

As a final remark, the noninteracting limit ($K=1$) of our results can also be derived by the scattering function formalism as done for the chiral Majorana case.\cite{Law} For repulsive interactions ($K<1$), one can in principle use the Bethe ansatz scattering eigenstates\cite{Pankaj,sp2} and derive the tunneling current for a single point contact. This formulation might be an extension of the perturbative approach introduced here, had the issues of complex Bethe momenta be clarified.\cite{sp2} Different type of interacting leads realizing different kind of Luttinger liquids\cite{Simon} can also be connected with noninteracting Majorana modes, which leaves unique transport signature due to different scaling behavior for an ideal single point contact. 
 
\acknowledgments
 SPC acknowledges the support by Taiwan's MOST (No.103-2811-M-001-112), the NCTS, the summer school support from ICAM to Weihai, Shandong, China, and the support by the Simons Foundation for the stay at Aspen, C.O., U.S.A., where part of this work is done. TLS is supported by the National Research Fund, Luxembourg (ATTRACT 7556175). CHC is supported by NSC grant No.98-2918-I-009-06, No.98-2112-M-009-010-MY3, the NCTU-CTS,
the MOE-ATU program, and the NCTS of Taiwan, R.O.C.

\appendix
\section{Cutoff dependence of differential conductance}\label{A0}
\begin{figure}
\includegraphics[width=.9\columnwidth]{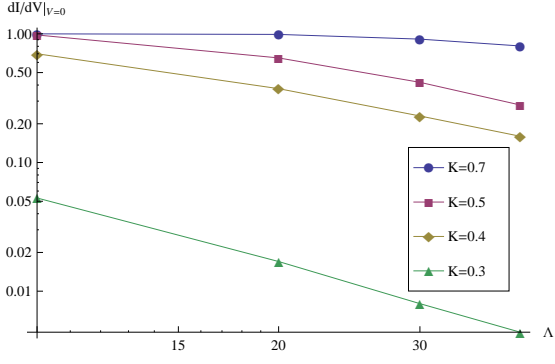}
\caption{Zero bias differential conductance v.s. linear momentum cutoff $\Lambda$ for fixed tunneling amplitude $\bar{t}$.}
 \label{scaledep}
\end{figure}

In this section we evaluate numerically the zero bias differential conductance at various different cutoffs, maintaining the cutoff energies at the order of $10^{-2}\epsilon_F$. In the main text we choose $\Lambda=10$ (i.e. with $\epsilon_F/\hbar v_F=10^3$ inverse length unit) and tunneling amplitude $\bar{t}=0.5$ in Fig. \ref{f3} and Fig. \ref{f4} and different $\bar{t}$ for the rest of the figures. Here, for demonstration purpose, we \emph{fix} $\bar{t}=0.5$ and all other parameters the same as those in Fig. \ref{f3} and Fig. \ref{f4}, and we vary $\Lambda$ from $10$ to $40$ stepped by $10$. The result is shown in log-log scale in the Fig. \ref{scaledep}. For $K_{critical}\simeq0.26<K<1$ the lowest order RG flow to the fixed point of perfect transmission, giving $dI/dV|_{V\rightarrow 0}=1$ (in unit of $g_1e^2/h$). 

In Fig. \ref{f3}  the calculated zero bias differential conductance depends explicitly on the linear momentum cutoff $\Lambda$, with larger $\Lambda$ giving smaller value and a generic trend of decreasing transmission amplitude for decreasing $K$ (or stronger repulsion in helical Luttinger lead, shown for $K=0.7$ to $0.3$). For generic two loops RG (or higher order) the next leading correction normally takes the form of $(1+1/F(\Lambda))$, with $1/F(\Lambda)\rightarrow 0$ as $\Lambda\rightarrow 0$. The explicit form of the higher order corrections $F(\Lambda)$ depends on the specific Hamiltonian. The trend we see in Fig. \ref{scaledep} is consistent with the naive higher order RG. However we shall bear in mind that this is not the RG type of calculations, but a fixed cutoff with inclusion of most of the perturbative terms (neglecting the level crossing terms) via the Dyson's approach. 
   
\section{Luttinger lead correlators}\label{AA}
The action 
\begin{eqnarray*} 
&&-S_0=\int_0^\beta d\tau \int dx \Big\{ [i\nabla\Theta(x,\tau)\partial_{\tau}\Phi(x,\tau)\\
&&-\frac{v}{2}(K(\nabla\Theta)^2+\frac{1}{K}(\nabla\Phi)^2 )]+\sum_\sigma \gamma_\sigma(\partial_\tau-\epsilon_d)\gamma_\sigma\Big\}
\end{eqnarray*}
At zero temperature
\begin{eqnarray}\nonumber
&&\frac{1}{K}\langle \Phi(r_1)\Phi(r_2)\rangle=\frac{-1}{2\pi}\ln\left[\frac{x^2+(a+iv t)^2}{a^2}\right]\\
&&  \equiv F^{(1)-+}(t,x)\\\nonumber
&&\langle \Phi(r_1)\Theta(r_2)\rangle=\frac{-1}{2\pi}\ln\left[\frac{a+iv t-i x}{a+iv t+i x}\right]\\
&&\equiv F^{(2)-+}(t,x)
\end{eqnarray}
Here $t=t_1-t_2$ and $x=x_1-x_2$.
For $t_2$ on the bottom and $t_1$ on the top branch of Keldysh contour we substitute $x\rightarrow -x$ and $t_1\leftrightarrow t_2$ to get
\begin{eqnarray}
&&F^{(1)+-}(t,x)=\frac{-1}{2\pi}\ln\left[\frac{x^2+(a-iv t)^2}{a_0^2}\right]\\
&&F^{(2)+-}(t,x)=\frac{-1}{2\pi}\ln\left[\frac{a-iv t+i x}{a_0-iv t-i x}\right]
\end{eqnarray}
For both $t_2$ and $t_1$ on the top branch, or time ordered branch, we get
\begin{eqnarray}\nonumber
F^{(1)++}(t,x)&=&\theta(t)F^{(1)-+}(t,x)+\theta(-t)F^{(1)+-}(t,x)\\\label{a6}
&=&\frac{-1}{2\pi}\ln\left[\frac{x^2+(a+iv |t|)^2}{a^2}\right]\\\nonumber
F^{(2)++}(t,x)&=&\theta(t)F^{(2)-+}(t,x)+\theta(-t)F^{(2)+-}(t,x)\\\label{a7}
&=&\frac{-1}{2\pi}\ln\left[\frac{a+iv |t|-i sgn[t] x}{a+iv |t|+i sgn[t]x}\right]
\end{eqnarray}
Similarly for anti-time ordered $F^{(1)--}(t,x)$ and $F^{(2)--}(t,x)$, obtained by $\theta(t)\leftrightarrow\theta(-t)$ in Eq.(\ref{a6}) and Eq.(\ref{a7}), are
\begin{eqnarray}
F^{(1)--}(t,x)&=&\frac{-1}{2\pi}\ln\left[\frac{x^2+(a-iv|t|)^2}{a^2}\right]\\
F^{(2)--}(t,x)&=&\frac{-1}{2\pi}\ln\left[\frac{a-iv |t|+i sgn[t] x}{a-iv |t|-i sgn[t]x}\right]
\end{eqnarray}
We absorb the effect of Klein factor $-i\langle T_c \eta_{R/L}(\tau_1)\eta_{R/L}(\tau_2)\rangle$ by introducing $\tilde{F}^{(2)++/--}(t,x)=F^{(2)++/--}(t,x)+F^{(2)++/--}(-t,-x)\pm sgn[t]i$ and $\tilde{F}^{(2)+-/-+}(t,x)=F^{(2)+-/-+}(t,x)+F^{(2)+-/-+}(-t,-x)\pm i$.
The general form of $G_{\psi_{j}}(\omega,x)$ at zero temperature is
\begin{eqnarray}\label{ggr}
&&G_{\psi_{j}}(\omega,x)=\frac{e^{isgn[j]k_F x}}{2\pi a}\\\nonumber
&&\times\int_{-\infty}^{\infty}dt e^{i(\omega-\mu)t}e^{\frac{\pi}{2}\left(\left(K+\frac{1}{K}\right)F^{(1)}(t,x)+sgn(j)\tilde{F}^{(2)}(t,x)\right)}
\end{eqnarray}
It is straightforward to show that the $G_{\psi_{j}}^{+-,-+}(t,x)$ obtained in Eq.(\ref{ggr}) (before performing the Fourier transform to frequency space) is the same as Eq.(A1) in Ref.~\onlinecite{Simon}. 
\begin{eqnarray}\nonumber
&&G_{\psi_{j}}^{\frac{+-}{-+}}(t,x)=\frac{e^{isgn[j]k_F x}}{2\pi a}\left[\frac{a}{sgn[j]x+v(t\pm i0^+)}\right]^{\kappa-\frac{1}{2}}\\
&&\times\left[\frac{a}{sgn[j]x-v(t\pm i0^+)}\right]^{\kappa+\frac{1}{2}}\\\nonumber
&&G_{\psi_{j}}^{\frac{++}{--}}(t,x)=\frac{e^{isgn[j]k_F x}}{2\pi a}\left[\frac{a}{sgn[j]x+v(t\mp isgn[t]0^+)}\right]^{\kappa-\frac{1}{2}}\\
&&\times\left[\frac{a}{sgn[j]x-v(t\mp isgn[t]0^+)}\right]^{\kappa+\frac{1}{2}}
\end{eqnarray}
 To compute Eq.(\ref{ggr}) let us first define $I^{\frac{+-}{-+}}(\omega,x)$ and $I^{\frac{++}{--}}(\omega,x)$ as
\begin{eqnarray}\label{ap9}
&&I^{\frac{+-}{-+}}(\omega,x)=\int_{-\infty}^{\infty}dt e^{i\omega t}\left[\frac{a^2}{x^2-v^2(t\pm i0^+)^2}\right]^{\kappa-\frac{1}{2}}\\\nonumber
&&I^{\frac{++}{--}}(\omega,x)=\int_{-\infty}^{\infty}dt e^{i\omega t}\left[\frac{a^2}{x^2-v^2(t\mp isgn[t]0^+)^2}\right]^{\kappa-\frac{1}{2}}
\end{eqnarray}
From Eq.(\ref{ap9}) it is easy to check that Eq.(\ref{ggr}) is expressed as
\begin{eqnarray}\nonumber
G_{\psi_{j}}^{\frac{+-}{-+}}(\omega,x)=\frac{-e^{i sgn[j]k_F x}}{4\pi (\kappa-\frac{1}{2})}\left[\partial_{sgn[j]x}+i\frac{\omega-\mu}{v}\right]I^{\frac{+-}{-+}}(\omega-\mu,x)\\\nonumber
G_{\psi_{j}}^{\frac{++}{--}}(\omega,x)=\frac{-e^{i sgn[j]k_F x}}{4\pi (\kappa-\frac{1}{2})}\left[\partial_{sgn[j]x}+i\frac{\omega-\mu}{v}\right]I^{\frac{++}{--}}(\omega-\mu,x)
\end{eqnarray}
Since $\omega$ is real we have $(I^{+-}(-\omega,x))^{\ast}=I^{-+}(\omega,x)$ and $(I^{++}(-\omega,x))^{\ast}=I^{--}(\omega,x)$. We only need to evaluate $I^{++}$ and $I^{+-}$.
For $I^{+-}(\omega,x)$ the nonzero contribution comes from the lower half circular contour.  
\begin{eqnarray}\nonumber
&&I^{+-}(\omega,x)=\left(\frac{a}{v}\right)^{2\kappa-1}\int_{-\infty}^{\infty}dt \frac{e^{i\omega t}}{\left(\left(\frac{x}{v}\right)^2-(t+i0^+)^2\right)^{\kappa-\frac{1}{2}}}\\\nonumber
&&=\left(\frac{a}{v}\right)^{2\kappa-1}\bigg\{-e^{i\omega\frac{|x|}{v}}\int_{C_1}\frac{e^{i\omega y} dy}{[-y(y+2|x|/v)]^{\kappa-\frac{1}{2}}}\\\nonumber
&&-e^{i\omega\frac{-|x|}{v}}\int_{C_2}\frac{e^{i\omega \bar{y}} d\bar{y}}{[\bar{y}(-\bar{y}+2|x|/v)]^{\kappa-\frac{1}{2}}}\bigg\}\theta(-\omega)\\\nonumber
&&=i\frac{2a\sqrt{\pi}}{v\Gamma(\kappa-\frac{1}{2})}\bigg\{\left(\frac{2i|x|v}{\omega a^2}\right)^{1-\kappa}K_{\kappa-1}(|x|\omega/iv)\\\label{ap13}
&&-\left(\frac{-2i|x|v}{\omega a^2}\right)^{1-\kappa}K_{\kappa-1}(-|x|\omega/iv)\bigg\}\theta(-\omega)
\end{eqnarray}

Here $K_n(z)$ is the modified Bessel function of the second kind and $\Gamma(x)$ is the Gamma function. For evaluation of $I^{++}(\omega,x)$, notice that $I^{++}(\omega,x)$ is an even function of $\omega$ following its definition:
\begin{eqnarray}
&&I^{++}(\omega,x)\\\nonumber
&&=\left(\frac{a}{v}\right)^{2\kappa-1}\int_{-\infty}^{\infty}dt \frac{e^{i\omega t}}{\left(\left(\frac{x}{v}\right)^2-(t-isgn[t]0^+)^2\right)^{\kappa-\frac{1}{2}}}
\end{eqnarray}
Thus we only need to evaluate $\omega>0$ in $I^{++}(\omega,x)$. For this $\omega>0$ region we have
\begin{eqnarray}\nonumber
&&I^{++}(\omega,x)\theta(\omega)\\\nonumber
&&=\left(\frac{a}{v}\right)^{2\kappa-1}\Big\{\int_{0}^{\infty} dt e^{i\omega t}\left(\left(\frac{x}{v}\right)^2-(t-i0^+)^2\right)^{-\kappa+\frac{1}{2}} \\\nonumber
&&+\int_{-\infty}^{0} dt e^{i\omega t}\left(\left(\frac{x}{v}\right)^2-(t+i0^+)^2\right)^{-\kappa+\frac{1}{2}}\Big\}\theta(\omega)\\\nonumber
&&=\left(\frac{a}{v}\right)^{2\kappa-1}\Big\{\int_{0}^{\infty} dt e^{i\omega t}\left(\left(\frac{x}{v}\right)^2-(t-i0^+)^2\right)^{-\kappa+\frac{1}{2}} \\\nonumber
&&+\int_{0}^{\infty} dt e^{-i\omega t}\left(\left(\frac{x}{v}\right)^2-(t-i0^+)^2\right)^{-\kappa+\frac{1}{2}}\Big\}\theta(\omega)\\\nonumber
&&=\left(\frac{a}{v}\right)^{2\kappa-1}\Big\{\Big[-e^{i\omega\frac{|x|}{v}}\int_{C_3}\frac{e^{i\omega y} dy}{[-y(y+2|x|/v)]^{\kappa-\frac{1}{2}}}\\\nonumber
&&-\int_{\infty}^0\frac{ie^{-\omega y} dy}{\left[\left(\frac{x}{v}\right)^2+y^2\right]^{\kappa-\frac{1}{2}}}\Big]-\Big[\int_0^{\infty}\frac{ie^{-\omega y} dy}{\left[\left(\frac{x}{v}\right)^2+y^2\right]^{\kappa-\frac{1}{2}}}\Big]\Big\}\theta(\omega)\\
&&=i\frac{2a\sqrt{\pi}}{v\Gamma(\kappa-\frac{1}{2})}\left(\frac{2i|x|v}{\omega a^2}\right)^{1-\kappa}K_{\kappa-1}(|x|\omega/iv)\theta(\omega)
\end{eqnarray}
The full expression for $I^{++}(\omega,x)$ is
\begin{eqnarray}\nonumber
&&I^{++}(\omega,x)=i\frac{2a_0\sqrt{\pi}}{v\Gamma(\kappa-\frac{1}{2})}\Bigg\{\left(\frac{2i|x|v}{\omega a^2}\right)^{1-\kappa}K_{\kappa-1}(|x|\omega/iv)\\\label{ap14}
&&\times\theta(\omega)+\left(\frac{-2i|x|v}{\omega a^2}\right)^{1-\kappa}K_{\kappa-1}(-|x|\omega/iv)\theta(-\omega)\Bigg\}
\end{eqnarray}
\begin{figure}
\includegraphics[width=.5\columnwidth]{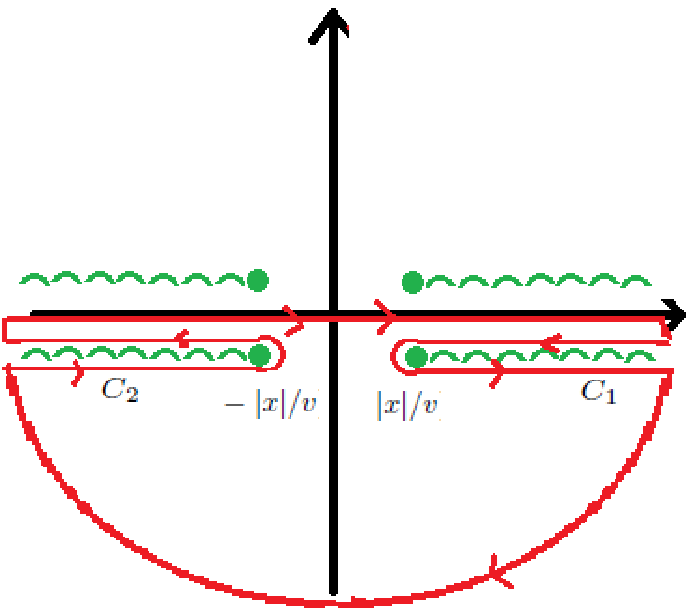}\hfill
\includegraphics[width=.5\columnwidth]{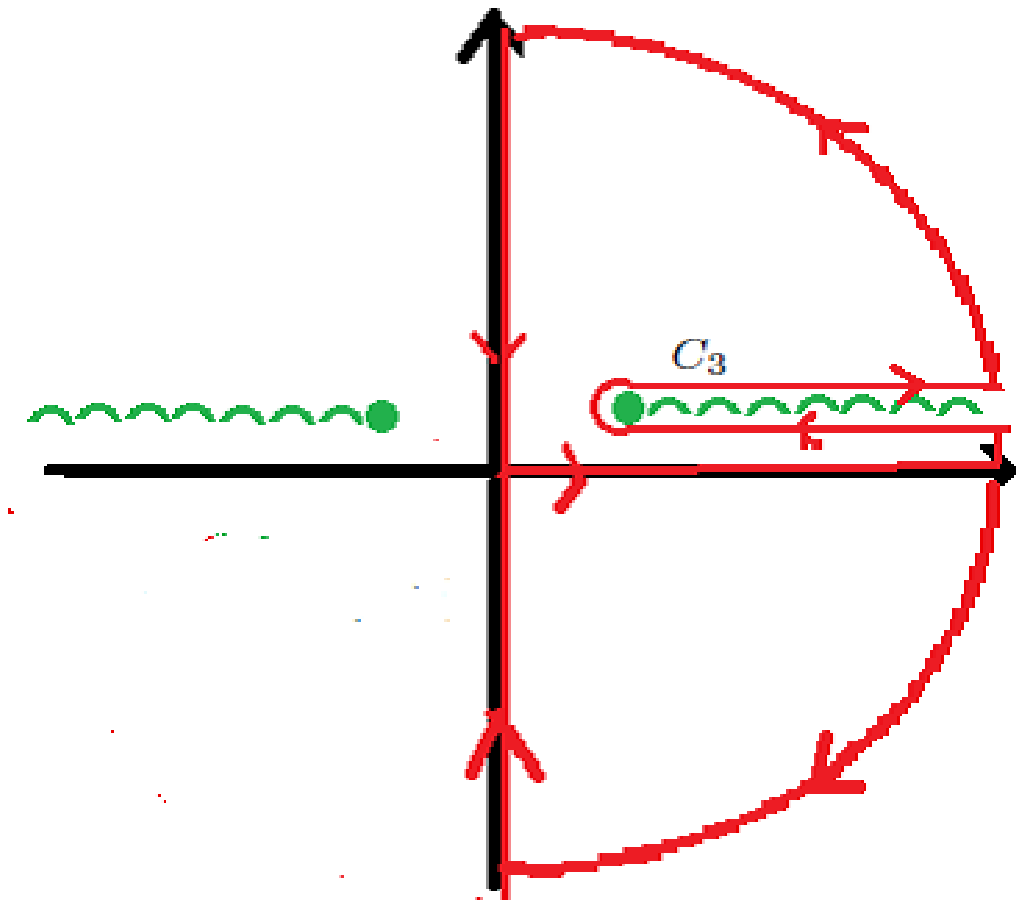}
\caption{Left: Contour chosen to evaluate $I^{+-}(\omega,x)$. Right: Contour chosen to evaluate $I^{++}(\omega,x)$}
 \label{f1a0}
\end{figure}
We combine the above results and use the derivative relation $$\partial_x\left(\frac{x}{a}\right)^{-n}K_n(ax)=-a\left(\frac{x}{a}\right)^{-n}K_{n+1}(ax)$$ for the modified Bessel functions. Replacing $\omega$ by $\omega-\mu$ to account for nonzero chemical potential and after some algebras we get
\begin{eqnarray}\nonumber
&&G_{\psi_{j}}^{+-}(\omega,x)=\frac{e^{isgn[j]k_F x}}{2\sqrt{\pi}\Gamma\left(\kappa+\frac{1}{2}\right)}\frac{a(\omega-\mu)}{v^2}\Bigg\{\left(\frac{2i|x|v}{(\omega-\mu)a^2}\right)^{1-\kappa}\\\nonumber
&&\times\left[K_{\kappa-1}\left(\frac{|x|(\omega-\mu)}{iv}\right)+sgn[jx]K_{\kappa}\left(\frac{|x|(\omega-\mu)}{iv}\right)\right]\\\nonumber
&&-\left(\frac{-2i|x|v}{(\omega-\mu)a^2}\right)^{1-\kappa}\bigg[K_{\kappa-1}\left(\frac{|x|(\omega-\mu)}{-iv}\right)-sgn[jx]\\\label{gpn}
&&\times K_{\kappa}\left(\frac{|x|(\omega-\mu)}{-iv}\right)\bigg]\Bigg\}\theta(\mu-\omega)
\end{eqnarray}
\begin{eqnarray}\nonumber
&&G_{\psi_{j}}^{-+}(\omega,x)=\frac{-e^{isgn[j]k_F x}}{2\sqrt{\pi}\Gamma\left(\kappa+\frac{1}{2}\right)}\frac{a(\omega-\mu)}{v^2}\Bigg\{\left(\frac{2i|x|v}{(\omega-\mu)a^2}\right)^{1-\kappa}\\\nonumber
&&\times\left[K_{\kappa-1}\left(\frac{|x|(\omega-\mu)}{iv}\right)+sgn[jx]K_{\kappa}\left(\frac{|x|(\omega-\mu)}{iv}\right)\right]\\\nonumber
&&-\left(\frac{-2i|x|v}{(\omega-\mu)a^2}\right)^{1-\kappa}\bigg[K_{\kappa-1}\left(\frac{|x|(\omega-\mu)}{-iv}\right)-sgn[jx]\\
&&\times K_{\kappa}\left(\frac{|x|(\omega-\mu)}{-iv}\right)\bigg]\Bigg\}\theta(\omega-\mu)
\end{eqnarray}
\begin{eqnarray}\nonumber
&&G_{\psi_{j}}^{++}(\omega,x)=\frac{e^{isgn[j]k_F x}}{2\sqrt{\pi}\Gamma\left(\kappa+\frac{1}{2}\right)}\frac{a(\omega-\mu)}{v^2}\Bigg\{\left(\frac{2i|x|v}{(\omega-\mu)a^2}\right)^{1-\kappa}\\\nonumber
&&\times\Bigg[K_{\kappa-1}\left(\frac{|x|(\omega-\mu)}{iv}\right)+sgn[jx]K_{\kappa}\left(\frac{|x|(\omega-\mu)}{iv}\right)\Bigg]\\\nonumber
&&\times\theta(\omega-\mu)+(\text{same expressions with } |x|\rightarrow -|x|\\
&&\text{and } sgn[jx]\rightarrow sgn[-jx])\theta(\mu-\omega)\Bigg\}
\end{eqnarray}
\begin{eqnarray}\nonumber
&&G_{\psi_{j}}^{--}(\omega,x)=\frac{-e^{isgn[j]k_F x}}{2\sqrt{\pi}\Gamma\left(\kappa+\frac{1}{2}\right)}\frac{a(\omega-\mu)}{v^2}\Bigg\{\left(\frac{2i|x|v}{(\omega-\mu)a^2}\right)^{1-\kappa}\\\nonumber
&&\times\Bigg[K_{\kappa-1}\left(\frac{|x|(\omega-\mu)}{iv}\right)+sgn[jx]K_{\kappa}\left(\frac{|x|(\omega-\mu)}{iv}\right)\Bigg]\\\nonumber
&&\times\theta(\mu-\omega)+(\text{same expressions with } |x|\rightarrow -|x| \\\label{gnn}
&&\text{and } sgn[jx]\rightarrow sgn[-jx])\theta(\omega-\mu)\Bigg\}
\end{eqnarray}
The $x\rightarrow 0$ limit is obtained by noting that the small argument expansion of the modified Bessel function of the second kind $K_{\alpha}(z)$ takes the following form\cite{mbook}:
\begin{eqnarray}\label{A21}
K_{\alpha}(z)\simeq\frac{1}{2}\left[\Gamma(\alpha)(\frac{2}{z})^\alpha+\Gamma(-\alpha)(\frac{z}{2})^\alpha\right](1+O(z^2))
\end{eqnarray}
for $z\rightarrow 0$. For $\alpha>0$ the first term in Eq.(\ref{A21}) is divergent, reflecting the artifact of the lack of small distance cutoff in taking the continuous limit (i.e. the smallest distance we can take should not be $x=0$ but lattice constant $a\propto 1/\Lambda$). Thus these divergence terms can be safely 
neglected or suppressed by the regularization via a further differentiation\cite{Simon}. Using further the functional relation of Gamma function $\Gamma(2z)=(2\pi)^{-1/2}2^{2z-\frac{1}{2}}\Gamma(z)\Gamma(z+\frac{1}{2})$ we can get Eq.(\ref{gsingle}) from Eq.(\ref{gpn}) to Eq.(\ref{gnn}) which is consistent with the direct derivation done in Ref.~\onlinecite{sp}.

\section{Derivation for Green functions}\label{AB}
Here we use the Langreth rule\cite{Langreth} on the Eq.(\ref{eq15}) and Eq.(\ref{eq16}). The retarded and advanced Green's functions are decoupled from lesser and greater ones and are solved directly from these two coupled equations. The explicit expressions of all retarded Green's functions expressed via unperturbed Majorana Green's functions and lead Luttinger Green's functions are:  
\begin{eqnarray}
&&G_{\gamma_\alpha}^R=\frac{G_{\gamma_\alpha}^{(0)R}-|t_\beta|^2 G_{\psi_\beta}^{R} (G_{\gamma_\alpha}^{(0)R}G_{\gamma_\beta}^{(0)R}-G_{\gamma_{\alpha}\gamma_{\beta}}^{(0)R}G_{\gamma_{\beta}\gamma_{\alpha}}^{(0)R})}{f_{\gamma}^R}\\\nonumber
&&G_{\gamma_{\alpha}\gamma_{\beta}}^R=\frac{G_{\gamma_{\alpha}\gamma_{\beta}}^{(0)R}+t_\alpha t_\beta^{\ast}G_{\psi_{\alpha} \psi_{\beta}}^{R}(G_{\gamma_\alpha}^{(0)R}G_{\gamma_\beta}^{(0)R}-G_{\gamma_{\alpha}\gamma_{\beta}}^{(0)R}G_{\gamma_{\beta}\gamma_{\alpha}}^{(0)R})}{f_{\gamma}^R}
\end{eqnarray}
\begin{eqnarray}
&&G_{\gamma_\beta}^R=\frac{G_{\gamma_\beta}^{(0)R}-|t_\alpha|^2 G_{\psi_\alpha}^{R} (G_{\gamma_\alpha}^{(0)R}G_{\gamma_\beta}^{(0)R}-G_{\gamma_{\alpha}\gamma_{\beta}}^{(0)R}G_{\gamma_{\beta}\gamma_{\alpha}}^{(0)R})}{f_{\gamma}^R}\\\nonumber
&&G_{\gamma_{\beta}\gamma_{\alpha}}^R=\frac{G_{\gamma_{\beta}\gamma_{\alpha}}^{(0)R}+t_\alpha^{\ast} t_\beta G_{\psi_{\beta}\psi_{\alpha} }^{R}(G_{\gamma_\alpha}^{(0)R}G_{\gamma_\beta}^{(0)R}-G_{\gamma_{\alpha}\gamma_{\beta}}^{(0)R}G_{\gamma_{\beta}\gamma_{\alpha}}^{(0)R})}{f_{\gamma}^R}
\end{eqnarray}
Here the numerator $f_{\gamma}^R$ is defined as 
\begin{widetext}
\begin{eqnarray}\nonumber
f_{\gamma}^R&\equiv& 1-G_{\gamma_\alpha}^{(0)R}|t_\alpha|^2 G_{\psi_\alpha}^{R}-t_\beta t^{\ast}_\alpha G_{\gamma_{\alpha}\gamma_{\beta}}^{(0)R}G_{\psi_{\beta} \psi_{\alpha}}^{R}-|t_\beta|^2 G_{\gamma_\beta}^{(0)R}G_{\psi_\beta}^{R}-t_\alpha t_\beta^{\ast}G_{\gamma_{\beta} \gamma_{\alpha}}^{(0)R}G_{\psi_{\alpha} \psi_{\beta}}^{R}+|t_\alpha|^2|t_\beta|^2 \\&\times&(G_{\gamma_\alpha}^{(0)R}G_{\gamma_\beta}^{(0)R}-G_{\gamma_\alpha \gamma_\beta}^{(0)R}G_{\gamma_\alpha \gamma_\beta}^{(0)R})(G_{\psi_{\alpha}}^R G_{\psi_{\beta}}^R-G_{\psi_{\alpha}\psi_{\beta}}^R G_{\psi_{\beta}\psi_{\alpha}}^R)
\end{eqnarray}
\end{widetext}
Once we obtain the full advanced and retarded Green's functions we then substitute these expressions into the equations for lesser and greater ones, which are coupled with advanced and retarded Green's functions. 
The full expressions for lesser Green's functions are
\begin{widetext}
\begin{eqnarray*}
G_{\gamma_\alpha}^{<}&=&\frac{Nu_{\alpha 1}}{De_1 De_2}-\frac{Nu_{\alpha 2}}{De_2}\\
Nu_{\alpha 1}&=&(1-t_\alpha t_\beta^\ast G_{\psi_{\alpha}\psi_{\beta}}^{R}G_{\gamma_\beta\gamma_\alpha}^{(0)R}-|t_\beta|^2 G_{\psi_{\beta}}^{R}G_{\gamma_\beta}^{(0)R})\Big((1-|t_\alpha|^2 G_{\psi_\alpha}^{R}G_{\gamma_\alpha}^{(0)R}-t_\alpha^\ast t_\beta G_{\psi_\beta \psi_\alpha}^R G_{\gamma_\alpha \gamma_\beta}^{(0)R})\\
&\times&\{G_{\gamma_\beta\gamma_\alpha}^{(0)<}+G_{\gamma_\beta}^A [ |t_\alpha|^2 (G_{\gamma_\beta\gamma_\alpha}^{(0)<}G_{\psi_\alpha}^A+G_{\gamma_\beta\gamma_\alpha}^{(0)R}G_{\psi_\alpha}^<)+t_\alpha^\ast t_\beta (G_{\gamma_\beta}^{(0)<}G_{\psi_\beta \psi_\alpha}^A+G_{\gamma_\beta}^{(0)R}G_{\psi_\beta \psi_\alpha}^< )]\\
&+&G_{\gamma_\beta \gamma_\alpha}^A[|t_\beta|^2 (G_{\gamma_\beta}^{(0)<}G_{\psi_\beta}^A+G_{\gamma_\beta}^{(0)R}G_{\psi_\beta}^< )+t_\alpha t_\beta^\ast (G_{\gamma_\beta\gamma_\alpha}^{(0)<}G_{\psi_\alpha \psi_\beta}^A+G_{\gamma_\beta\gamma_\alpha}^{(0)R}G_{\psi_\alpha \psi_\beta}^<)]\}\\
&+&(|t_\alpha|^2 G_{\psi_\alpha}^R G_{\gamma_\beta\gamma_\alpha}^{(0)R}+t_\alpha^\ast t_\beta G_{\psi_\beta \psi_\alpha}^R G_{\gamma_\beta}^{(0)R})\{
G_{\gamma_\alpha}^{(0)<}+G_{\gamma_\alpha}^A [|t_\alpha|^2 (G_{\gamma_\alpha}^{(0)<} G_{\psi_\alpha}^A +G_{\gamma_\alpha}^{(0)R} G_{\psi_\alpha}^<)\\
&+&t_\alpha^\ast t_\beta (G_{\gamma_\alpha \gamma_\beta}^{(0)<}G_{\psi_\beta \psi_\alpha}^A+G_{\gamma_\alpha \gamma_\beta}^{(0)R}G_{\psi_\beta \psi_\alpha}^<)]+G_{\gamma_\beta\gamma_\alpha}^A [|t_\beta|^2 (G_{\gamma_\alpha \gamma_\beta}^{(0)<}G_{\psi_\beta}^A+G_{\gamma_\alpha \gamma_\beta}^{(0)R}G_{\psi_\beta}^<)\\
&+&t_\alpha t_\beta^\ast (G_{\gamma_\alpha}^{(0)<}G_{\psi_{\alpha}\psi_\beta}^A+G_{\gamma_\alpha}^{(0)R}G_{\psi_{\alpha}\psi_\beta}^<)]\}\Big)\\
Nu_{\alpha 2}&=&G_{\gamma_\beta\gamma_\alpha}^{(0)<}+|t_\alpha|^2(G_{\gamma_\beta\gamma_\alpha}^{(0)<}G_{\psi_\alpha}^A G_{\gamma_\alpha}^A+G_{\gamma_\beta\gamma_\alpha}^{(0)R}G_{\psi_\alpha}^< G_{\gamma_\alpha}^A)+t_\alpha^\ast t_\beta(G_{\gamma_\beta}^{(0)<}G_{\psi_\beta \psi_\alpha}^A G_{\gamma_\alpha}^A +G_{\gamma_\beta}^{(0)R}G_{\psi_\beta \psi_\alpha}^< G_{\gamma_\alpha}^A )\\
&+&t_\alpha t_\beta^\ast (G_{\gamma_\beta \gamma_\alpha}^{(0)<}G_{\psi_\alpha \psi_\beta}^A G_{\gamma_\beta \gamma_\alpha}^{A}+G_{\gamma_\beta \gamma_\alpha}^{(0)R}G_{\psi_\alpha \psi_\beta}^< G_{\gamma_\beta \gamma_\alpha}^{A})+|t_\beta|^2(G_{\gamma_\beta}^{(0)<}G_{\psi_\beta}^A G_{\gamma_\beta \gamma_\alpha}^A +G_{\gamma_\beta}^{(0)R}G_{\psi_\beta}^< G_{\gamma_\beta \gamma_\alpha}^A )\\
De_1&=&(1-|t_\alpha|^2 G_{\psi_{\alpha}}^{R}G_{\gamma_{\alpha}}^{(0)R}-t_\alpha^\ast t_\beta G_{\psi_\beta\psi_{\alpha}}^R G_{\gamma_{\alpha}\gamma_{\beta}}^{(0)R})(1-|t_\beta|^2 G_{\psi_{\beta}}^{R}G_{\gamma_{\beta}}^{(0)R}-t_\alpha t_\beta^\ast G_{\psi_\alpha \psi_\beta}^R G_{\gamma_\beta \gamma_\alpha}^{(0)R})\\
&-&(|t_\alpha|^2 G_{\psi_{\alpha}}^{R}G_{\gamma_{\beta}\gamma_{\alpha}}^{(0)R}+t_\alpha^\ast t_\beta G_{\psi_\beta\psi_{\alpha}}^R G_{\gamma_{\beta}}^{(0)R})(|t_\beta|^2 G_{\psi_{\beta}}^{R}G_{\gamma_{\alpha}\gamma_{\beta}}^{(0)R}+t_\alpha t_\beta^\ast G_{\psi_{\alpha}\psi_\beta}^R G_{\gamma_\alpha}^{(0)R})\\
De_2&=&|t_\alpha|^2 G_{\psi_{\alpha}}^{R} G_{\gamma_{\beta}\gamma_{\alpha}}^{(0)R}+t_\alpha^\ast t_\beta G_{\psi_\beta\psi_{\alpha}}^R G_{\gamma_{\beta}}^{(0)R}
\end{eqnarray*}
\end{widetext}
\begin{widetext}
\begin{eqnarray*}
G_{\gamma_\beta \gamma_\alpha}^{<}&=&\frac{Nu_{\beta\alpha}}{De_1}\\
Nu_{\beta\alpha}&=&(1-|t_\alpha|^2 G_{\psi_\alpha}^R G_{\gamma_\alpha}^{(0)R}-t_\alpha^\ast t_\beta G_{\psi_\beta \psi_\alpha}^R G_{\gamma_\alpha \gamma_\beta}^{(0)R})\{G_{\gamma_\beta \gamma_\alpha}^{(0)<}+G_{\gamma_\alpha}^A [|t_\alpha|^2 (G_{\gamma_\beta \gamma_\alpha}^{(0)<} G_{\psi_\alpha}^A
+G_{\gamma_\beta \gamma_\alpha}^{(0)R} G_{\psi_\alpha}^<)\\
&+&t_\alpha^\ast t_\beta(G_{\gamma_\beta}^{(0)<}G_{\psi_\beta \psi_\alpha}^A+G_{\gamma_\beta}^{(0)R}G_{\psi_\beta \psi_\alpha}^<)]+G_{\gamma_\beta \gamma_\alpha}^A[t_\alpha t_\beta^\ast (G_{\gamma_\beta \gamma_\alpha}^{(0)<}G_{\psi_\alpha \psi_\beta}^A+G_{\gamma_\beta \gamma_\alpha}^{(0)R}G_{\psi_\alpha \psi_\beta}^<)\\
&+&|t_\beta|^2(G_{\gamma_\beta}^{(0)<}G_{\psi_\beta}^A+G_{\gamma_\beta}^{(0)R}G_{\psi_\beta}^<)]\}+(|t_\alpha|^2 G_{\psi_\alpha}^R G_{\gamma_\beta \gamma_\alpha}^{(0)R}+t_\alpha^\ast t_\beta G_{\psi_\beta \psi_\alpha}^R G_{\gamma_\beta}^{(0)R})\{G_{\gamma_\alpha}^{(0)<}+G_{\gamma_\alpha}^A\\
&\times&[|t_\alpha|^2(G_{\gamma_\alpha}^{(0)<}G_{\psi_\alpha}^A+G_{\gamma_\alpha}^{(0)R}G_{\psi_\alpha}^<)+t_\alpha^\ast t_\beta (G_{\gamma_\alpha \gamma_\beta}^{(0)<}G_{\psi_\beta \psi_\alpha}^A+G_{\gamma_\alpha \gamma_\beta}^{(0)R}G_{\psi_\beta \psi_\alpha}^<)]+G_{\gamma_\beta \gamma_\alpha }^A[|t_\beta|^2\\
&\times&(G_{\gamma_\alpha \gamma_\beta}^{(0)<}G_{\psi_\beta}^A+G_{\gamma_\alpha \gamma_\beta}^{(0)R}G_{\psi_\beta}^<)+t_\alpha t_\beta^\ast (G_{\gamma_\alpha}^{(0)<}G_{\psi_\alpha \psi_\beta}^A+G_{\gamma_\alpha}^{(0)R}G_{\psi_\alpha \psi_\beta}^<)]\}
\end{eqnarray*}
\end{widetext}


\begin{thebibliography}{10}
\bibitem{Hasan}M. Z. Hasan and C. L. Kane, Rev. Mod. Phys. {\bf 82}, 3045 (2010).
\bibitem{Qi}X. Qi and S.-C. Zhang, Rev. Mod. Phys. {\bf 83}, 1057 (2011).
\bibitem{Essin}A. M. Essin and V. Gurarie, Phys. Rev. B {\bf 84}, 125132 (2011).
\bibitem{Mong}R. S. K. Mong and V. Shivamoggi, Phys. Rev. B {\bf 83}, 125109 (2011).
\bibitem{Bernevig}B. A. Bernevig, T. L. Hughes, and S.-C. Zhang, Science \textbf{314}, 1757 (2006).
\bibitem{KaneMele}C. L. Kane and E. J. Mele, Phys. Rev. Lett. \textbf{95} 226801 (2005).
\bibitem{Konig}M. K{\"o}nig, S. Wiedmann, C. Br{\"u}ne, A. Roth, H. Buhmann, L. W. Molenkamp, X.-L. Qi and S.-C. Zhang, Science {\bf 318}, 766 (2007).
\bibitem{Knez}I. Knez, R.-R. Du, and G. Sullivan, Phys. Rev. Lett. {\bf 107}, 136603 (2011).
\bibitem{CWJ}C. W. J. Beenakker, Annu. Rev. Con. Mat. Phys. {\bf 4}, 113 (2013). 
\bibitem{Alicea}J. Alicea, Rep. Prog. Phys. {\bf 75}, 076501 (2012).
\bibitem{Mourik}V. Mourik, K. Zuo, S. M. Frolov, S. R. Plissard, E. P. A. M. Bakkers, and L. P. Kouwenhoven, Science {\bf 336}, 1003 (2012)
\bibitem{NadjPerge}S. Nadj-Perge, I. K. Drozdov, J. Li, H. Chen, S. Jeon, J. Seo, A. H. MacDonald, B. A. Bernevig, and A. Yazdani, Science {\bf 346}, 602 (2014)
\bibitem{FuKane}L. Fu, and C. L. Kane, Phys. Rev. Lett. {\bf 100}, 096407 (2008).
\bibitem{XLQi}X.-L. Qi, T. L. Hughes, S. Raghu, and S.-C. Zhang, Phys. Rev. Lett. {\bf 102}, 187001 (2009). 
\bibitem{Tanaka}Y. Tanaka, T. Yokoyama, A. V. Balatsky, and N. Nagaosa, Phys. Rev. B {\bf 79}, 060505 (2009).
\bibitem{Sato}M. Sato and S. Fujimoto, Phys. Rev. B {\bf 79}, 094504 (2009). 
\bibitem{Wang}J. Wang, Y. Xu, and S.-C. Zhang, Phys. Rev. B {\bf 90}, 054503 (2014).
\bibitem{Queiroz}R. Queiroz and A. P. Schnyder, Phys. Rev. B {\bf 91}, 014202 (2015).
\bibitem{Wu}C. Wu, B. A. Bernevig, and S.-C. Zhang, Phys. Rev. Lett. {\bf 96}, 106401 (2006).
\bibitem{Xu}C. Xu, J. E. Moore, Phys. Rev. B {\bf 73}, 045322 (2006).
\bibitem{Du}T. Li, X. Mu, X. Liu, P. Wang, H. Fu, X. Lin, K. Schreiber, G. Csathy, L. Du, G. Sullivan, R.-R. Du, 2015 APS march meeting, unpublished.    
\bibitem{Liu}C.-X. Liu and B. Trauzettel, Phys. Rev. B {\bf 83}, 220510(R) (2011).
\bibitem{Beri}B. B$\acute{e}$ri, Phys. Rev. B {\bf 85}, 140501(R) (2012).
\bibitem{Nagaosa}Y. Asano, Y. Tanaka, and N. Nagaosa, Phys. Rev. Lett. {\bf 105}, 056402 (2010).
\bibitem{Loss}S. Gangadharaiah, B. Braunecker, P. Simon, and D. Loss, Phys. Rev. Lett. {\bf 107}, 036801 (2011).
\bibitem{Orth15}C. P. Orth, R. P. Tiwari, T. Meng, and T. L. Schmidt, arXiv:1405.4353 [cond-mat.mes-hall] (2014)
\bibitem{Fidkowski}L. Fidkowski, J. Alicea, N. H. Lindner, R. M. Lutchyn, and M. P. A. Fisher, Phys. Rev. B {\bf 85}, 245121 (2012). 
\bibitem{YW}Y.-W. Lee and Y.-L. Lee, Phys. Rev. B {\bf 89}, 125417 (2014).
\bibitem{Affleck}I. Affleck and D. Giuliano, J. Stat. Mech., P06011 (2013).
\bibitem{Law}K. T. Law, P. A. Lee, and T. K. Ng, Phys. Rev. Lett. {\bf 103}, 237001 (2009).
\bibitem{Law2}C. L. M. Wong and K. T. Law, Phys. Rev. B {\bf 86}, 184516 (2012).
\bibitem{realhm}R. Queiroz, A. P. Schnyder, Phys. Rev. B. {\bf 91}, 014202 (2015).
\bibitem{Li}J. Li, G. Fleury, and M. B\"uttiker, Phys. Rev. B {\bf 85}, 125440 (2012).
\bibitem{TG}T. Giamarchi, Quantum physics in one dimension, Oxford university press (2004). 
\bibitem{Schmidt}T. L. Schmidt, Phys. Rev. Lett. {\bf 107}, 096602 (2011).
\bibitem{sp} S. P. Chao, S. A. Silotri, and C. H. Chung, Phys. Rev. B {\bf 88}, 085109 (2013).
\bibitem{footnote2} Extra factor of $2\pi$ in the denominator here comes from different definition of Fourier transform to frequency domain used in Eq.(A9) here and Eq.(A10) in Ref.~\onlinecite{sp}.
\bibitem{shHo} S. H. Ho, S. P. Chao, C. H. Chou, and F. L. Lin, New J. Phys. {\bf 16}, 113062 (2014).
\bibitem{Martin}D. Chevallier, J. Rech, T. Jonckheere, C. Wahl, and T. Martin, Phys. Rev. B {\bf 82}, 155318 (2010).
\bibitem{Sassetti}G. Dolcetto, S. Barbarino, D. Ferraro, N. Magnoli, and M. Sassetti, Phys. Rev. B {\bf 85}, 195138 (2012).
\bibitem{footnote} Overall magnitude here means we sum the peaks' magnitude if one peak splits into two due to interference. 
\bibitem{Zhang}C. X. Liu, H. J. Zhang, B. Yan, X. L. Qi, T. Frauenheim, X. Dai, Z. Fang, and S. C. Zhang, Phys. Rev. B {\bf 81}, 041307 (2010).
\bibitem{Shen}H. Z. Lu, W. Y. Shan, W. Yao, Q. Niu, and S. Q. Shen, Phys. Rev. B {\bf 81}, 115407 (2010). 
\bibitem{Wen}C. de C. Chamon, D. E. Freed, S. A. Kivelson, S. L. Sondhi, and X. G. Wen, Phys. Rev. B {\bf 55}, 2331 (1997).
\bibitem{Rosenow}B. I. Halperin, A. Stern, I. Neder, and B. Rosenow, Phys. Rev. B {\bf 83}, 155440 (2011).
\bibitem{Stern}D. E. Feldman, Y. Gefen, A. Kitaev, K. T. Law, and A. Stern, Phys. Rev. B {\bf 76}, 085333 (2007).
\bibitem{Slingerland}P. Bonderson, K. Shtengel, and J. K. Slingerland, Annals of Physics {\bf 323}, 2709 (2008).
\bibitem{Slingerland2}P. Bonderson, K. Shtengel, and J. K. Slingerland, Phys. Rev. Lett. {\bf 97} 016401 (2006).
\bibitem{Beenakker}A. R. Akhmerov, Johan Nilsson, and C. W. J. Beenakker, Phys. Rev. Lett. {\bf 102}, 216404 (2009).
\bibitem{Recher}P. Virtanen and P. Recher, Phys. Rev. B {\bf 83}, 115332 (2011). 
\bibitem{Rizzo}B. Rizzo, L. Arrachea, and M. Moskalets, Phys. Rev. B {\bf 88}, 155433 (2013).
\bibitem{Huang} C. W. Huang, S. T. Carr, D. Gutman, E. Shimshoni, A. D. Mirlin, Phys. Rev. B {\bf 88}, 125134 (2013).
\bibitem{Dolcini}F. Dolcini, Phys. Rev. B {\bf 83}, 165304 (2011).
\bibitem{Orth13}C. P. Orth, G. Str\"ubi, T. L. Schmidt, Phys. Rev. B, {\bf 88}, 165315 (2013).
\bibitem{Natan}R. Citro, F. Romeo, and N. Andrei, Phys. Rev. B {\bf 84}, 161301 (2011).
\bibitem{Pankaj}P. Mehta and N. Andrei, Phys. Rev. Lett. {\bf 96}, 216802 (2006).
\bibitem{sp2}S. P. Chao and G. Palacios, Phys. Rev. B {\bf 83}, 195314 (2011).
\bibitem{mbook}M. Abramowitz and I. A. Stegun, Handbook of Mathematical Functions, Applied Mathematics Series, Vol. 55 (National Bureau
of Standards, Washington, DC, 1972).
\bibitem{Simon}B. Braunecker, C. Bena, and P. Simon, Phys. Rev. B {\bf 85}, 035136 (2012). 
\bibitem{Langreth}  H. Haug and A.-P. Jauho, Quantum Kinetics in transport and optics of semiconductors, Springer-Verlag (1998).
\end{thebibliography}
\end{document}